\documentclass{aa}  
\usepackage{graphicx}
\usepackage{txfonts}

\begin{document}

   \title{C and N abundances in globular clusters. I. The case of 47 Tuc and the effect of the first dredge-up.}

   \subtitle{Implications for the isochrone fitting}

   \titlerunning{C and N abundances and isochrone fitting in 47 Tuc.}

   \authorrunning{Villanova et al. 2024}

   \author{S. Villanova
          \inst{1}
          \and
          L. Monaco
          \inst{1,2}
          \and
          Y. Momany
          \inst{3}
          \and
          A. Plotnikova
          \inst{4}
          \and
          I. Ordenes
          \inst{5}
          }

   \institute{Universidad Andres Bello, Facultad de Ciencias Exactas, Departamento de F{\'i}sica y Astronom{\'i}a - Instituto de Astrof{\'i}sica, Autopista
Concepci\'on-Talcahuano 7100, Talcahuano, Chile
             \email{sandro.villanova@unab.cl}
        \and
            INAF-Osservatorio Astronomico di Trieste, Via G.B. Tiepolo 11,34143 Trieste, Italy
        \and
             Osservatorio Astronomico di Padova-INAF, Vicolo dell'Osservatorio 5, I-35122 Padova, Italy\\
             \email{yazan.almomany@inaf.it}
        \and 
            Dipartimento di Fisica e Astronomia, Università di Padova, I-35122, Padova, Italy\\
            \email{anastasiia.plotnikova@studenti.unipd.it}
        \and
            Departamento de Astronomía, Casilla 160-C, Universidad de Concepción, Concepción 4030000, Chile
             }

   \date{Received ......; accepted ......}

  \abstract 
   {Globular clusters exhibit  star-to-star chemical variations, traceable through both photometric and  spectroscopic data.  While UV photometry and light-element abundances  (e.g., Na and O) are  commonly used, the optical V  vs. (V-I)  color-magnitude diagram (CMD)  is often assumed to  be relatively  unaffected by  such inhomogeneities  and is typically used  to derive  basic cluster  parameters.  However C and N are the best tracers  of these variations but are  challenging to measure  due to their spectral features lying in the blue/UV or IR regions.}
   {In  this study,  we investigate  chemical variations  in the  globular cluster  NGC104  (47Tucanae),  aiming   to  trace  multiple  stellar
populations  across  evolutionary  phases  and  examine  how  the  C/N anti-correlation evolves from the main sequence (MS) to the asymptotic
giant branch (AGB). We also assess  the impact of these populations on the interpretation of the V  vs. V-I diagram.}
   {Using spectra spanning  all evolutionary stages, we  derive [C/Fe] and [N/Fe]  abundances  for  a  large  stellar  sample.   These  abundance measurements  are inferred  from the  CN  and the  CH features,  while atmospheric parameters are homogeneously derived from photometry.  The inferred abundances allow us to disentangle multiple populations along the CMD and refine cluster parameters.}
   {We find that MS  stars are more C- and N-rich than  their red giant branch, horizontal branch, and AGB counterparts. The C/N  anti-correlation shifts during the sub-giant branch phase,  coinciding with the first  dredge-up: C decreases by  0.15-0.20 dex,  N  by $\sim$0.1 dex,  while  Fe  remains unchanged.  Interestingly, stars  with  different C  and N  abundances occupy distinct regions of the V vs V-I diagram--a pattern not  attributable  to  differential  reddening.   Proper  CMD  fitting requires two  isochrones with  differing helium  content, metallicity, and possibly age.}
   {}


   \maketitle

\section{Introduction}

Galactic globular clusters (GCs) are known to host star-to-star variations of their chemical content. More specifically, \cite{car09} showed that all Galactic GCs have at least a spread (or anti-correlation) in the content of their light-elements O and Na. In some cases also a Mg and Al spread is observed \citep{pan17}. The only confirmed exception is Ruprecht 106, where \citet{vil13} found that stars show no variation. This spread is due to the early evolution of each cluster, formed initially by a first generation of stars that has the same chemical composition of field stars at the same metallicity. The subsequent generation of stars (Na-richer and O-poorer) is formed from gas polluted by ejecta of evolved stars of the first generation (Na-poorer and O-richer). This is the so-called multiple-population (MP) phenomenon. This spectroscopic evidence has been interpreted as the signature of material processed during H-burning at high temperatures by proton-capture reactions (such as the Ne–Na and Mg–Al cycles). Several polluters have been proposed: intermediate-mass asymptotic giant branch stars  \citep{dan16}, fast-rotating massive stars \citep{dec07}, and massive binaries \citep{dem09}. An attractive alternative has been proposed by \citet{bas13} that implies only a single burst of star formation. They postulate that the gas ejected from massive stars of the first (and only) generation concentrates in the center of the cluster and is acquired by low-mass stars via disc accretion while they are in the fully convective phase of the pre-main sequence. However none of the proposed scenarios can account for all observational constrains, as explained by \citet{ren15}.

\begin{figure}
   \centering
   \includegraphics[width=\hsize]{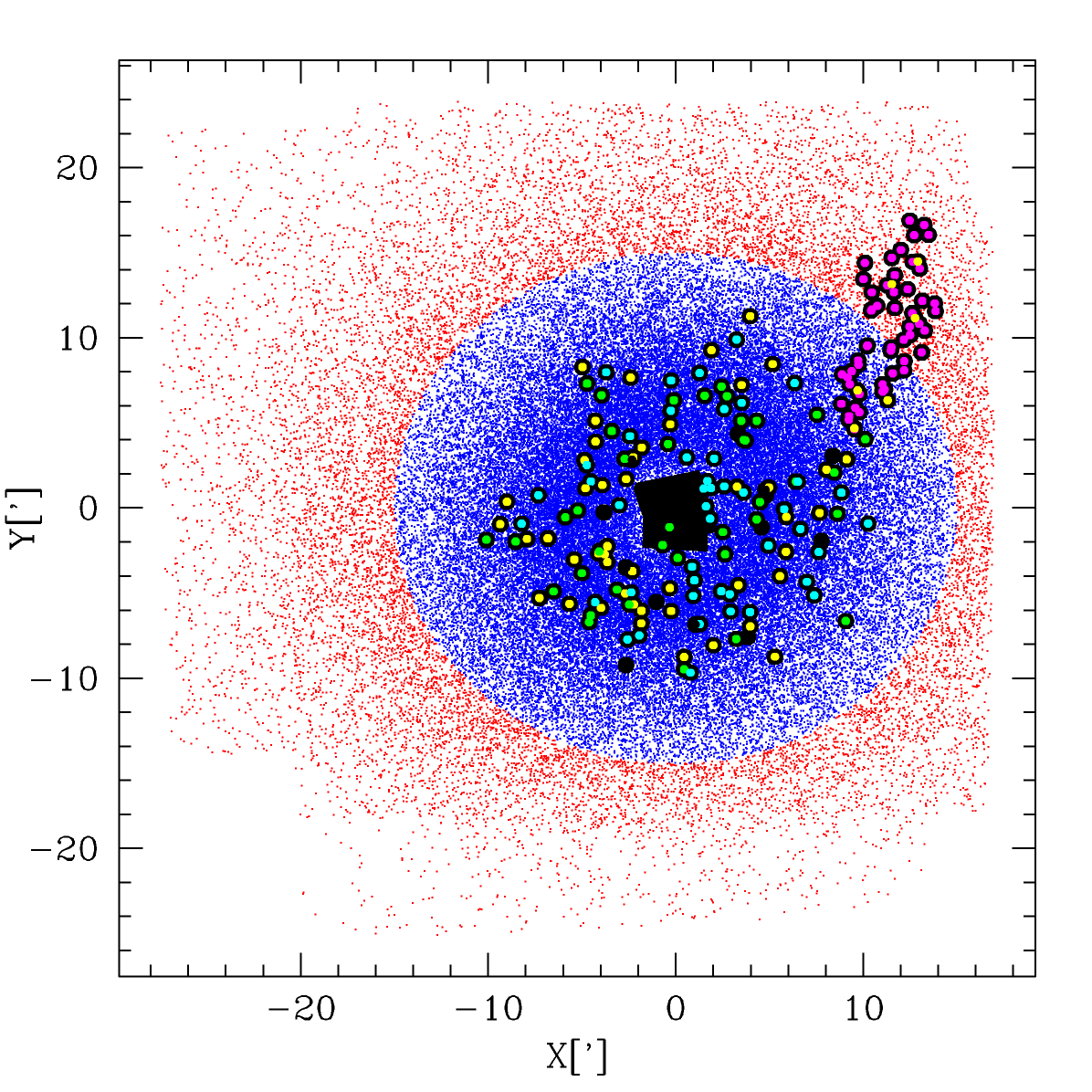}
   \caption{Distribution of NGC 104 member stars in the sky. Blue and red points are stars from ground-based photometry, 
   while black points in the center are stars from the  HST Large Legacy Treasury Program. The radial cut at 15’  was used to 
   separate the inner from the outer cluster region. Black circles with colored insets represent MS (magenta), SGB (yellow), RGB (green), HB (cyan) and AGB (black) targets. See text for more details. }
    \label{fig1}
    \end{figure}

In addition to the abundance spread of light elements, variations in heavier elements have also been found in some massive GCs, such as $\omega$ Centauri \citep{vil14}, M54 \citep{car10}, M22 \citep{mar09}, Terzan5 \citep{mas14} and NGC 2419 \citep{coh10}. However, these are generally thought to be the vestige of more massive primitive dwarf galaxies that merged with the Galaxy.

Parallel to the spectroscopic evidence, although found much later, an unexpected multiplicity in their color-magnitude diagram (CMD) sequences was discovered (e.g. \citealt{pio07} for NGC2808, \citealt{mil08} for NGC1851, and \citealt{and09} for 47Tuc). Dedicated mid-/near-ultraviolet Hubble Space Telescope surveys \citep{pio15,mil17,nar18} ultimately proved that multiplicity in the main sequence, the sub-giant branch and the red giant branch is the general rule rather than an exception. 

\begin{table}
\caption{Summary of the photometric databases we used in the paper. }          
\label{table0A}      
\centering                          
\begin{tabular}{l c c c}        
\hline\hline                 
Type   & Coverage & Data used in the paper       \\    
\hline                        
Ground-based & 2'-25'   & V,I filters   \\
HST          & 0'-3'    & F606W,F814W filters  \\
GAIA         & 0'-25'   & Proper motions\\
\hline                                   
\end{tabular}
\tablefoot{The second column report the radial coverage of the database and the third the data we used from each database.}
\end{table}

\begin{table*}
\caption{Summary of the spectroscopic databases we used in the paper.}         
\label{table0B}      
\centering                          
\begin{tabular}{l c c c c c c c}        
\hline\hline                 
Type         & Resolution &  Phases     & S/N     & Spectral coverage & N. of targets & C-range & N-range\\    
\hline                        
FORS         & 815        &  MS/SGB     &  50-90  & 3800-5500 & 58 & 4280-4315&3845-3885\\
GIRAFFE-LR02 & 6000       &  RGB/HB/AGB & 300-350 & 3964-4567 &  86& 4300-4315&4214-4216\\
GIRAFFE-HR04 & 24000      &  SGB/RGB    &  20-60  & 4188-4392 & 58 & 4300-4315&4214-4216\\
\hline                                   
\end{tabular}
\tablefoot{The second column reports the spectral resolution, the third the parts of the CMD covered by the spectra, the fourth the S/N range of the data, the fifth the full spectral range (in $\AA$), the sixth the number of good targets for each dataset, the seventh and eighth the spectral range we used to measure C and N abundances respectively (in $\AA$).}
\end{table*}

From what we just said, we can conclude that optical Na-O abundances and UV photometry have been the key observable to disentangle MP in GCs, at least until infrared spectroscopy arrived on to market with, as an example, the APOGEE \citep{maj17} and CAPOS \citep{gei21} surveys. These surveys allow to measure carbon and nitrogen abundances, for which the access in the optical is forbidden most of the times for the two reasons. In metal poor stars, such as most of the targets in GCs, carbon can be measured only from the G-band at 4300 \AA. This spectral region is not included in most of the default setups that have been used to study MPs like the UVES-580nm or the GIRAFFE-HR11/HR13 at the FLAMES instrument of the VLT and dedicated observations must be planned. Also, good N abundance estimations can be done only in the blue-UV region (that is, the blue CN 4125 \AA or other even bluer CN bands) or in the near IR, beyond 8000 \AA. The second reason is that blue and ultraviolet (UV) spectra are heavily crowded making abundance determination difficult, even with the use of spectrosynthesis and the up-to-date molecular linelists. 

As the APOGEE and CAPOS surveys showed, C and N abundances are very powerful tools to study MPs since the N variation in GC stars is extreme. Most of the time this variation is larger than 1 dex, and this makes the separation of the sub-populations much easier compared to O or Na. However the use of C and N have two limitations. The first is that it must always be coupled with a determination of the O abundances for stars colder than 4500 K where, due to the molecular equilibrium, the three atoms form CO, OH, CN and CH molecules. This is easy in the IR, where several OH lines are present in the giant stars that are the targets of the two survey (we must notice however that OH lines vanish rapidly when stars become hotter than 4500$\div$5000 K), but in the optical it requires the observation of the only O line at 6300 \AA, which is very far away from the CN and CH bands. The second is that surface C and N abundances are supposed to change moving from the MS to the red-giant branch (RGB) due to the effect of the first dredge-up \citep{sal20}. This effect is poorly studied and it makes difficult to assess the original abundances of the sub-populations. 

   \begin{figure}
   \centering
   \includegraphics[width=\hsize]{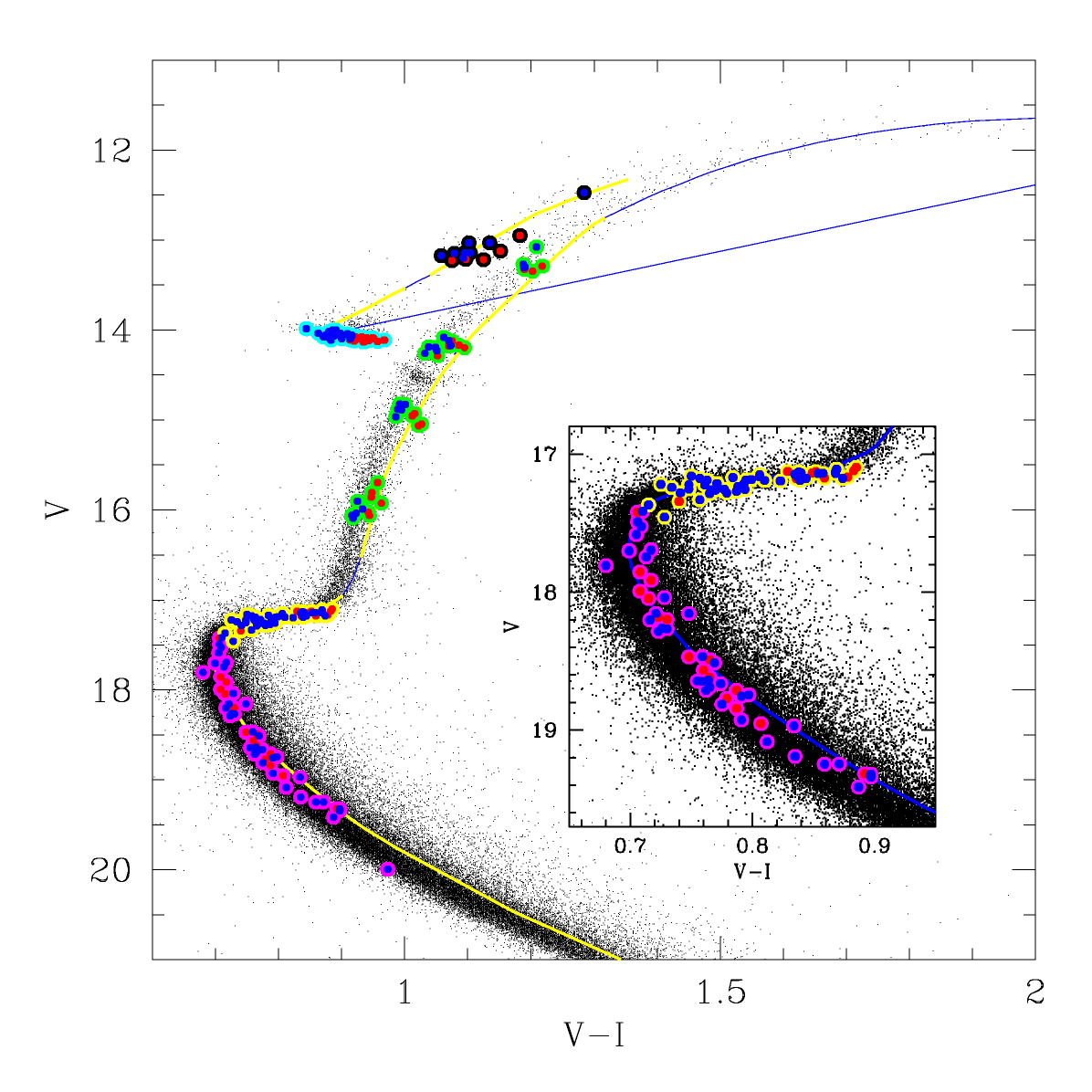}
   \caption{Position on the CMD of our targets. The blue/yellow line is the best fitting isochrone we used for the parameter determination. Blue and red symbols inside each circle represent FG and SG stars respectively. See text for more details.}
    \label{fig2}
    \end{figure}

NGC 104 or 47 Tuc is one of the most massive GCs in the Milky Way \citep{bau18} and was studied by several authors. The classical \citet{car09} paper found the typical Na-O anticorrelation, while \citet{nor82}, using low resolution spectra of a sample of HB stars, found the presence of a C vs. N anticorrelation with C-poor stars being redder on average and 0.04 fainter in their V magnitude. This result was later confirmed by \citet{bri97}. Finally \citet{mil12} showed that the cluster has a bimodal CMD where two populations can be easily be identified using UV filters. A minor third population (8\% of the total) can also be seen at the level of the SGB. 

   \begin{figure*}
   \centering
   \includegraphics[width=\hsize]{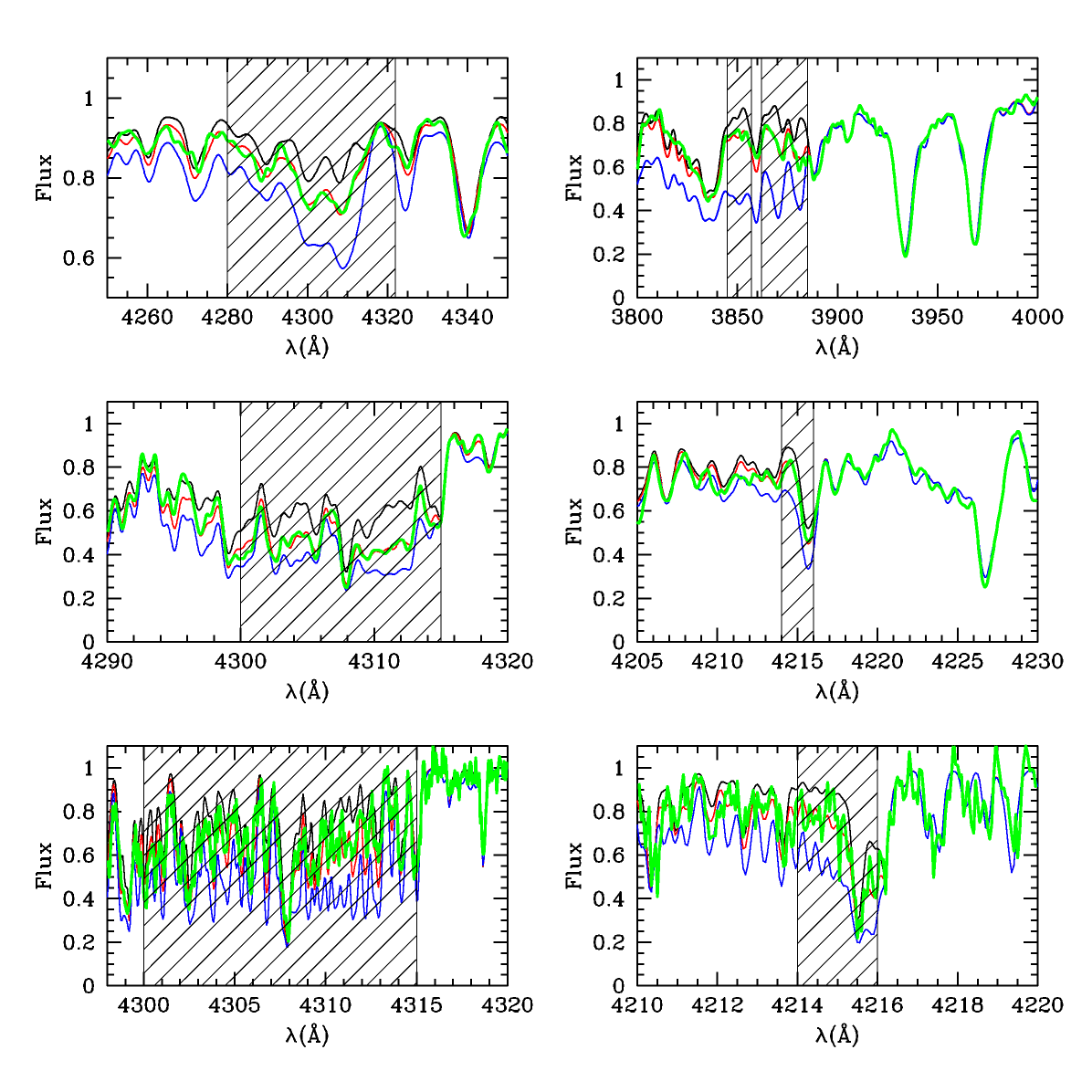}
   \caption{Example of spectrosynthesis. Panels on the left show the synthesis for C while panels on the right show the synthesis for N. Upper row shows the synthesis for the MS target \#1\_13, middle row shows synthesis for the HB target \#RHB\_I\_a\_40848, while the lower row shows the synthesis for the SGB target N104e\_51104. The wavelength range used for the abundance determination is shown as a shaded region. Observed spectra are in green. Red spectra show the best match, while blue and black spectra represent abundance variations of $\pm$0.5 dex for C respectively, while for N we used $\pm$1.0 dex for low resolution spectra of MS and SGB stars (upper panel) and $\pm$0.5 dex for intermediate and high resolution spectra of SGB,RGB,HB and AGB stars (middle and lower panel).}
    \label{fig3}
    \end{figure*}

In this paper we will use blue and UV spectroscopic data to trace the C and N abundances. These data cover the full CMD, from the main sequence (MS) to the asymptotic giant branch (AGB), so we will be able to trace the [C/Fe] and [N/Fe] variation in all the evolutionary phases, including the sub-giant branch (SGB), where the first dredge-up is suppose to happen. We have a collection of low, middle and high resolution spectra and high quality V vs. V-I photometry, that allows us to estimate homogeneous atmospheric parameters along the entire CMD. In this context, the most difficult parts to study are the MS and the turn-off (TO) because their targets are intrinsically faint but also because, for such hot stars (T$_{eff}\sim$6000 K), the CN feature at 4215 \AA, that can be used in the more advances phases, is no longer visible and the UV CN band at 3870 \AA\ must be considered. However this band (and the carbon G-band at 4300 \AA) is easily resolved in low resolution spectra, that also allow to obtain high S/N data for such faint targets.

Our investigation led us also to a deeper understanding of the V vs. V-I CMD. As mentioned earlier, UV photometry is used to separate MPs in GCs since the sub-population occupy different loci in UV CMDs. On the other hand the V vs. V-I or analogous CMDs have always been used to infer the basic parameters of the clusters such as the age, the distance and reddening \citep{mar09}, since FG and SG are generally assumed to be well mixed there. The only exception we know of is \citet{and09}, who found an intrinsic spread in the NGC 104 MS, 3 magnitudes below the TO and a split in the SGB region.  We will show that, thanks to our spectroscopic data, MPs can be disentangled in all the evolutionary phases also in the V vs. V-I CMD, and this has an important impact in the parameter determination of the cluster.

In section \ref{td} and \ref{fda} we will describe the data we used and the first spectroscopic analysis. Section \ref{FSG} shows how C and N abundances can be used to disentangle MPs from the MS to the AGB and how better parameters for the cluster can be obtained. This also allows to obtain better atmospheric parameters for the targets and in section \ref{sad} we will perform a second data analysis using these improved parameters. In section \ref{res} we will discuss the results, and we will give a summary in section \ref{con}.

\section{The data}  \label{td}

The data we used in this paper were obtained from different sources. 
As far as photometry is concerned, we used ground-based data provided by 
\citet{ste19} \footnote{https://www.cadc-ccda.hia-iha.nrc-cnrc.gc.ca/en/community/STETSON/} and HST data from the HST Large Legacy Treasury Program \citep{pio15,nar18} \footnote{https://groups.dfa.unipd.it/ESPG/treasury.php}. 
The ground-based photometry covers a radial range from $\sim$ 2’ to $\sim$ 25’ (see figure \ref{fig1})
in the B, V and I filters, and spans from the tip of the RGB to well below the main-sequence turnoff point. 
HST photometry is based on data obtained using the F275W, F336W, F438W, F606W, and F814W filters 
in the central 4’x4’ region (see figure \ref{fig1}) and spans from the tip of the RGB to several magnitudes below 
the main-sequence turnoff point. We complemented ground-based photometry with GAIA DR3 
(Gaia Collaboration et al. 2016B, Gaia Collaboration et al. 2023j)  \footnote{https://gea.esac.esa.int/archive/} in order to have proper 
motions and select member stars. A summary of the photometric data we used is reported in table \ref{table0A}.

   \begin{figure}
   \centering
   \includegraphics[width=\hsize]{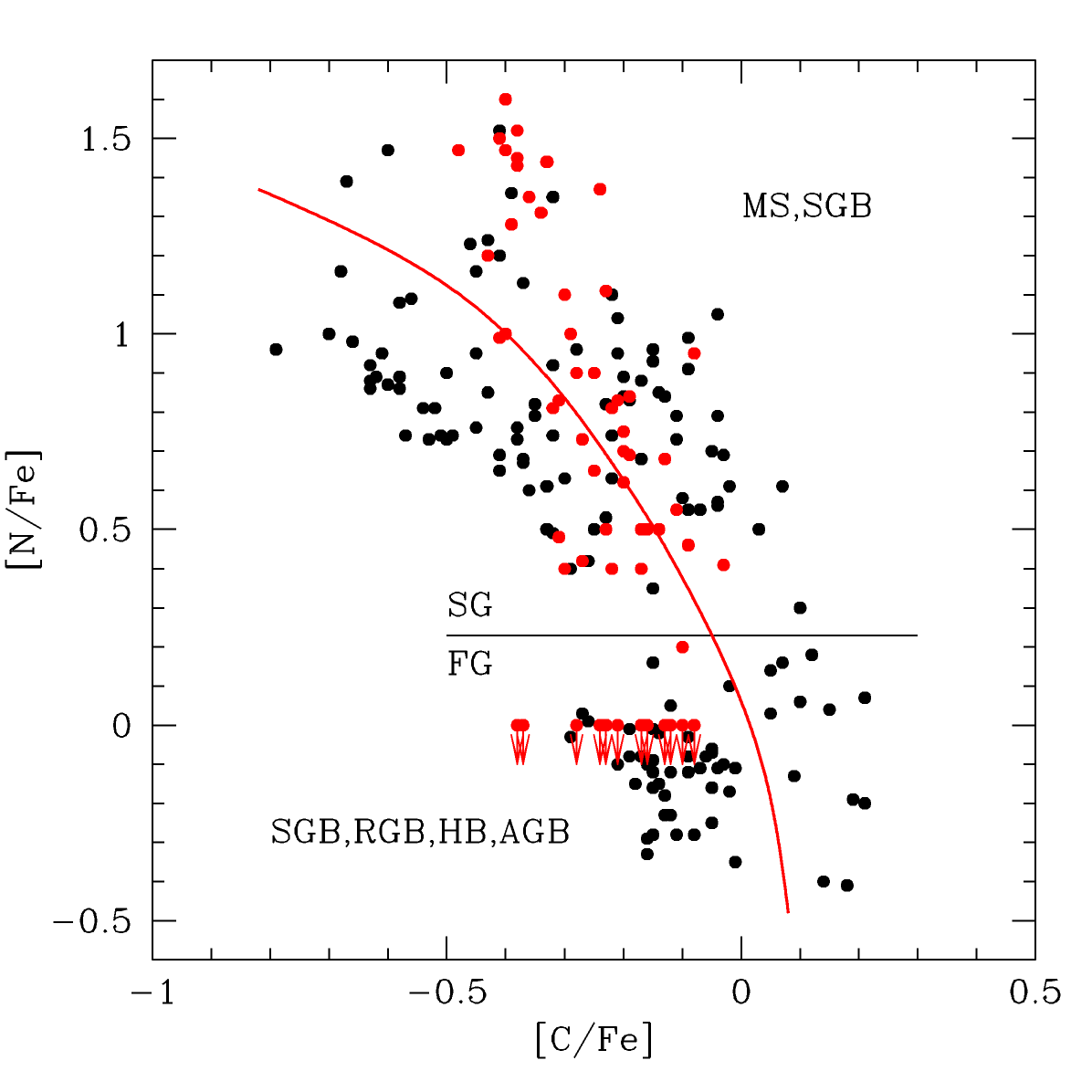}
   \caption{[N/Fe] vs. [C/Fe] anticorrelation as obtained from the first abundance determination run. The red line divides MS targets on the right from RGB/HB/AGB targets on the left. SGB targets are located on both sides of the red line. The black line divides FG from SG stars. Red arrows indicate stars for which \citet{mar16} gives no [N/Fe]. For these stars we assumed [N/Fe]=0.0 as an upper limit. See text for more details.}
   \label{fig4}
   \end{figure}

Spectroscopic data for the MS/SGB region were kindly provided by Briley (private communication) and are those published and described 
in   \citet{bri04}, while data for the RGB/HB/AGB region were obtain by the ESO program 105.2094.001 (P.I. Momany). The MS/SGB spectra (for 115 targets in total)
were obtained at the ESO VLT telescope using multi-slit masks with the Focal Reducer and Low Dispersion Spectrograph 2 (FORS2) \footnote{https://www.eso.org/sci/facilities/paranal/instruments/fors/inst/mxu.html} and cover a range
from 3800 to 5500 \AA\ with a resolution of R=815. A significant fraction of these targets were rejected for the analysis because of the too low S/N or because they  turned out to be non-members, leaving us with 58 targets for which good abundances have been obtained. 97 RGB/HB/AGB targets
were instead observed using the FLAMES@VLT+GIRAFFE spectrograph \footnote{https://www.eso.org/sci/facilities/paranal/instruments/flames/inst/Giraffe.html} 
with the LR02 grism and cover a range from 3964 to 4567 \AA\ with a resolution of R=6000. The sky was clear, 
and the typical seeing was 0.8 arcsec. Thirty-two 45 min spectra were obtained for each target. The data were reduced using GIRAFFE 
pipeline 2.16.10 \footnote{https://www.eso.org/sci/facilities/paranal/instruments/flames/tools/drs.html}, which corrects spectra for bias and flat-field. 
Then spectra were extracted and a wavelength calibration solution is applied based on non-simultaneous calibration lamps. After that, each 
spectrum was corrected for its fiber transmission  coefficient, which is found from flat-field images by measuring, for each fiber, the average 
flux relative to a reference fiber. Using IRAF \footnote{https://iraf-community.github.io/} we applied a sky correction to each stellar spectrum by
subtracting the average of the 10 sky spectra that were observed simultaneously with the stars (same FLAMES plate). Finally,
all of them were normalized to the continuum, i.e., divided by a low-order polynomial that fits its continuum, and the 32 spectra for each 
star were combined. The resulting spectra have a dispersion of 0.2 $\AA$ per pixel and a typical S/N~300-350. Of the 97 targets, 11 were not considered for the abundance analysis because they have temperature below 4500 K, which would required an oxygen abundance estimation to have reliable C and N contents, as discussed above.

\begin{table*}
\caption{Parameters and [C/Fe] and [N/Fe] abundances for our targets.}         
\label{table1}      
\centering                          
\begin{tabular}{l c c c c c c c c c c c c c}        
\hline\hline                 
ID & RA & DEC & V & I & PM$_{RA}$ & PM$_{DEC}$ & T$_{eff}$ & log(g) & v$_{t}$ & Phase & Gen. & [C/Fe] & [N/Fe]\\    
\hline                        
3\_22 & 6.55150000 & -71.96975000 & 18.71 & 17.95 & 5.26 & -2.45 & 5722 & 4.50 & 0.95 &  MS & SG & -0.17 & 0.97\\
2\_12 & 6.67933334 & -71.93763889 & 19.41 & 18.53 & 4.63 & -2.61 & 5332 & 4.59 & 0.88 &  MS & SG & -0.32 & 0.76\\
1\_06 & 6.58404167 & -71.87013889 & 17.85 & 17.14 & 5.20 & -2.35 & 5964 & 4.29 & 1.05 &  MS & FG & 0.01 & -0.26\\
1\_13 & 6.56370833 & -71.84133333 & 17.74 & 17.03 & 5.20 & -2.50 & 5977 & 4.26 & 1.06 &  MS & SG & -0.28 & 0.85\\
1\_18 & 6.63979166 & -71.83630556 & 18.49 & 17.72 & 5.63 & -3.08 & 5772 & 4.46 & 0.97 &  MS & FG & -0.07 & 0.01\\
3\_13b & 6.53808333 & -72.00433333 & 18.47 & 17.71 & 4.41 & -1.90 & 5782 & 4.45 & 0.97 &  MS & SG & -0.14 & 0.82\\
\hline                                   
\end{tabular}
\tablefoot{RA and DEC are in degrees, proper motions in mas/yr, temperature in K and microturbulence in km/s. {\it Phase} indicates the evolutionary phase of the target and {\it Gen.} indicates if the target belong to the FG or to the FG. The full version of the table is available online at CDS.}
\end{table*}

\begin{table*}
\caption{Errors on the abundance determination.}             
\label{table1B}      
\centering                          
\begin{tabular}{l c c c c c c c c}        
\hline\hline                 
$\Delta$(Abundance) & $\Delta$ T$_{eff}$ & $\Delta$ $\log$g & $\Delta$v$_{t}$ & S/N & $\sigma_{Tot}^{Par}$ & $\Delta${[}C/Fe{]} & $\Delta${[}N/Fe{]} & $\sigma_{Tot}$ \\
          &        30 K        &      0.05        &   0.05 km/s     &     &  & & &\\
\hline                        
$\Delta${[}C/Fe{]}$_{MS}$    &  0.04 & 0.01  & 0.00  & 0.05 & 0.06 &  -  & 0.00& 0.06\\
$\Delta${[}N/Fe{]}$_{MS}$    &  0.07 & 0.02  & 0.00  & 0.06 & 0.09 &  0.07  & -& 0.11\\   
\hline
$\Delta${[}C/Fe{]}$_{SGB}$   &  0.03 & 0.00  & 0.01  & 0.04 & 0.05 &  -  & 0.00& 0.05\\
$\Delta${[}N/Fe{]}$_{SGB}$   &  0.05 & 0.00  & 0.00  & 0.06 & 0.08 &  0.05  & -& 0.09\\ 
\hline
$\Delta${[}C/Fe{]}$_{RGB}$   &  0.02 & 0.00  & 0.01  & 0.04 & 0.05 &  -  & 0.00& 0.05\\
$\Delta${[}N/Fe{]}$_{RGB}$   &  0.04 & 0.02  & 0.01  & 0.05 & 0.07 &  0.05  & -& 0.09\\ 
\hline
$\Delta${[}C/Fe{]}$_{HB}$    &  0.04 & 0.00  & 0.01  & 0.05 & 0.06 &  -  & 0.00& 0.06\\
$\Delta${[}N/Fe{]}$_{HB}$    &  0.06 & 0.01  & 0.01  & 0.07 & 0.09 &  0.05  & -& 0.10\\ 
\hline
$\Delta${[}C/Fe{]}$_{AGB}$   &  0.02 & 0.00  & 0.02  & 0.04 & 0.05 &  -  & 0.00& 0.05\\
$\Delta${[}N/Fe{]}$_{AGB}$   &  0.03 & 0.01  & 0.01  & 0.05 & 0.06 &  0.05  & -& 0.08\\ 
\hline                                   
\end{tabular}
\tablefoot{Columns 2 to 4 indicate the [C/Fe] and [N/Fe] abundance variation due to the variation
in temperature, gravity and microturbulence indicated for each parameter. Column 5 shows the error due to the S/N. Column 6 is the total error calculated as the root-square of the sum of the squares of the previous 4 columns, and represents the total error due to the uncertainty on stellar parameters and to the S/N. Column 7 is the [C/Fe] variation due to the uncertainty in [N/Fe] given by column 6, while column 8 is the [N/Fe] variation due to the uncertainty in [C/Fe] given by column 6. Finally column 9 gives the total error calculated as the root-square of the sum of the squares of the previous 3 columns. This is the final error that we uses in the text.}
\end{table*}

The spectroscopic dataset was complemented with high-resolution spectra of 58 targets in the SGB region from the ESO program 089.D-0579 (P.I. Marino). Also in this case data 
were obtained using the FLAMES at VLT+GIRAFFE spectrograph, but with the HR04 grism and cover a range from 4188 to 4392 \AA\ with a resolution of R=24000. The sky was clear, 
and the typical seeing was 0.8 arcsec. Eleven 45 min spectra were obtained for each target. The data were reduced using GIRAFFE 
pipeline 2.16.10 applying the same procedure described before. Resulting spectra have a dispersion of 0.05 $\AA$ per pixel and a typical S/N~20-60. A summary of the spectroscopic data we used is reported in table \ref{table0B}.

We used the FXCOR utility within 
IRAF to measure the radial velocity, adopting a synthetic spectrum as a template calculated with a temperature of 5000 K, a gravity of 3.0, and a 
metallicity of [Fe/H]=-0.70. All the GIRAFFE targets were selected in order to be members using Gaia database, and the membership was also 
confirmed by radial velocities, that all agree within a dispersion of $\sim$10 km/s around the mean that is -16 km/s, a value very close to 
that given by \citet{har10}, i.e. -18 km/s.

All the targets are reported in table \ref{table1} where we give the ID, the coordinates, the V and I magnitudes, proper motions, atmospheric parameters (T$_{\rm eff}$, log(g), v$_{\rm t}$), the evolutionary phase (MS,SGB,RGB,HB,AGB), the population the target belong to (FG or SG), and the carbon and nitrogen abundances (see the following sections for the determination of the atmospheric parameter, the phase, the population and the abundances).

\section{First data analysis} \label{fda}

Our targets are plotted in figure \ref{fig1} and \ref{fig2}. According to their position on the CMD, we could  identify 52 MS (magenta), 53 SGB (yellow), 
67 RGB (green), 90 HB (cyan) and 26 AGB (black) stars. Inside each symbol of figure \ref{fig2} there is a red or blue circle which identify if the target belongs to the first or second generation respectively. This identification will be discussed below, when C and N abundances will be available. 
We anticipate that reliable C and N abundances were obtained only for those RGB/AGB targets having Teff$>$4500 K. This is because for 
lower temperatures oxygen starts to play an important role in the molecular equilibrium, and a small change of 0.1 dex in its abundance can have an 
impact of up to several tenths of dex in [C/Fe] or [N/Fe]. Several RGB stars were rejected but we kept an AGB star with  Teff=4329 K  (a temperature not far away from our limit) for completeness.
In order to obtain the atmospheric parameters we decided to use the V vs. V-I CMD because the V-I color depends only on 
temperature and not on metallicity according to \citet{alo99}. 
Then we performed an isochrone fitting using the Padova database \citep{bre12} \footnote{http://stev.oapd.inaf.it/cgi-bin/cmd}.
We varied age and [M/H] in order to obtain the best fit of the TO and the slope of the upper RGB regions respectively, and varied the 
apparent distance modulus (m-M)$_V$ and the reddening E(V-I) in order to properly fit all the evolutionary phases of the CMD.
The fitting was obtained by minimizing the distance between the isochrone and the photometry. For the MS and RGB phases we minimized the color difference at a given magnitude, while for the SGB, HB and AGB we minimized the magnitude difference at a given color. We also applied a visual check in order to avoid bad fitting to the data.
We obtained the following preliminary parameters: (m-M)$_V$=13.29, E(V-I)=0.03, Age=14.0 Gyrs and [M/H]=-0.60.

We notice that, while the fitting of the TO, upper MS, SGB, upper RGB, HB and AGB is quite good, the isochrone is slightly 
redder then the lower MS and the lower RGB. The first natural explanation would be that the isochrone has some modelling 
issue in those regions of the CMD but, as we will show later, the correct explanation is that two isochrones with different Y and
[M/H] values are required to properly fit the NGC 104 V vs. V-I CMD. On the other hand, the fit we propose here is good enough 
for a first estimations of the parameters. According to the evolutionary phase we used two different but complementary method.
Temperature for SGB and HB targets was estimated from the Teff vs. V-I relation that can be obtained from the isochrone at that evolutionary 
phases. Gravity for HB targets was just assumed to be log(g)=2.47, since this is the value given by the isochone for unevolved HB stars, while for the SGB
we used the log(g) vs. V-I relation that can be obtained from the isochrone. For the MS, the RGB and the AGB instead it is 
possible to obtain Teff vs. V and log(g) vs. V relations from the isochrone that return parameters with a significantly lower internal error than 
the classical Teff vs. V-I method, both because the V magnitude has an error $\sqrt{2}$ times lower than the V-I color and because such 
relations are intrinsically less sensitive to any internal error. As we will show later, this second method return abundances 
that are only marginally better that the first for RGB and AGB targets, but that are significantly better for the MS, where
errors on magnitudes are large for the faintness of the stars. We will show also that this first guess of the parameters 
is affected by internal systematic errors that will be corrected using the two isochrone fitting we mentioned above. Portions
of the isochrone used for the parameter determination are indicated by yellow segments in figure \ref{fig2}.
Having temperature and gravity for each target, microturbulence can be obtained using the relation by \citet{mot20}.
The final parameters we need for the spectral analysis is metallicity [Fe/H], for which we assumed a value of 
-0.70 dex \citep{har10}.

\begin{figure}
   \centering
   \includegraphics[width=\hsize]{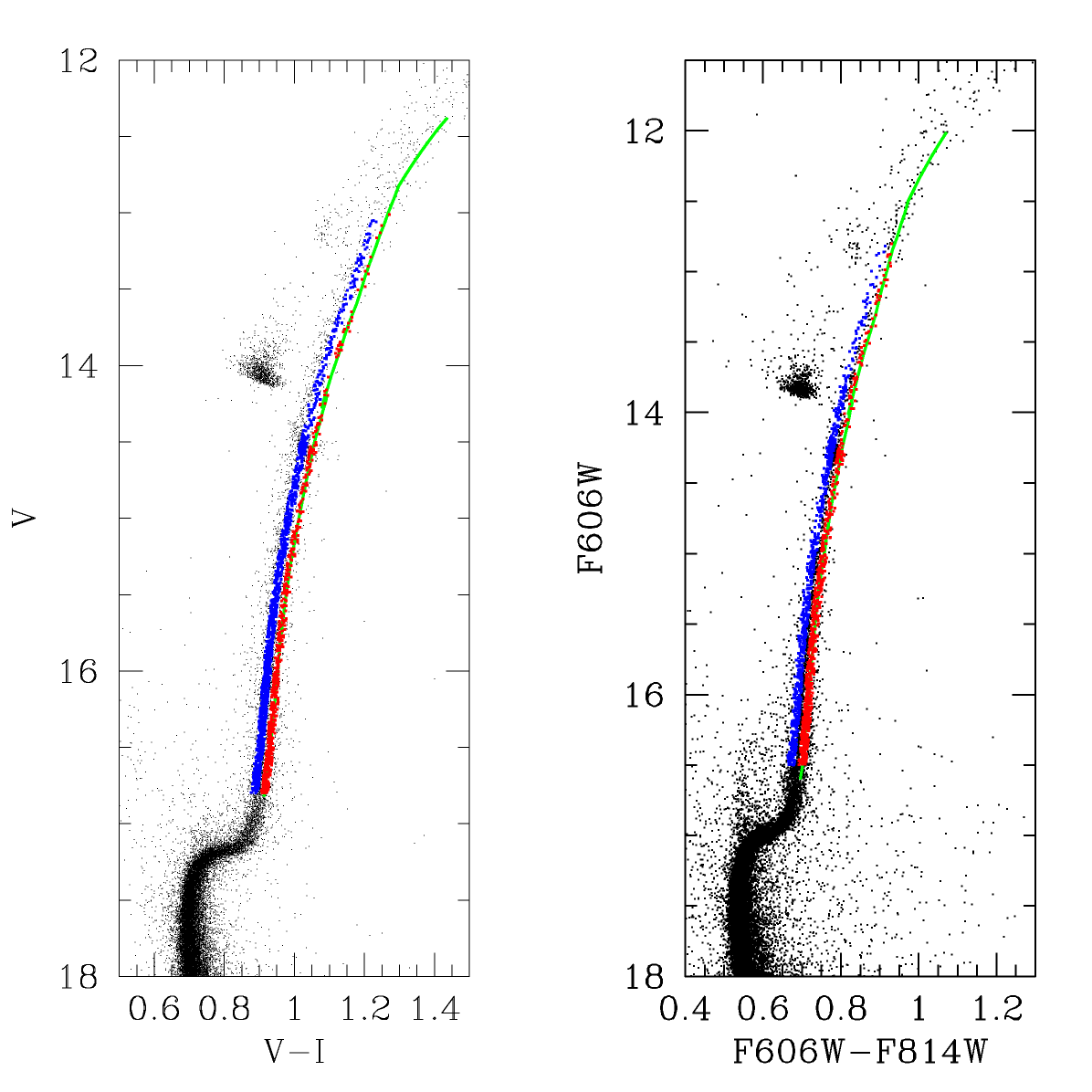}
   \caption{Ground-based (left) and HST (right) CMDs of the cluster. FG generation (red) and SG (blue) star selection is indicated. Green line is the isochrone used for this selection. See text for more details.}
   \label{fig5}
    \end{figure}

Atmospheric models were calculated using the ATLAS9 code \citep{kur70} 
assuming our initial estimations for atmospheric parameters. We then used the 2.77 version of the SPECTRUM 
software \footnote{https://www.appstate.edu/~grayro/spectrum/spectrum.html} to calculate synthetic spectra. We 
obtained C abundance from the G band. For low resolution spectra we used the range 4280-4315 \AA, while for the 
intermediate and high resolution spectra we used the 4300-4315 \AA range. For the N abundance instead, we used the CN band at 3845-3885 \AA\ for 
the low resolution spectra, while the CN band at 4214-4126 \AA\ for the intermediate and high resolution spectra. Examples of spectrosynthesis 
are reported in figure \ref{fig3}, while abundances are reported in figure \ref{fig4} and table \ref{table1}. For this first analysis we adopted the C and N abundances from \citet{mar16}
as far as HR04 data are concerned. We will review and refine these abundances during the second data analysis. We just notice here that \citet{mar16} gives no N content for a subsample of star, probably because the N feature was too weak to be measured. Since the lowest N content measured by \citet{mar16} is +0.2 dex, we give [N/Fe]=0.0 as upper limit to this subsample for the moment. We will show in section \ref{sad} that a reliable N content can be measure also for these targets. 

A clear [N/Fe] vs. [C/Fe] anticorrelation appears (see fig.\ref{fig4}) with unevolved MS targets having higher C and N abundances on average if compared with their more evolved counterparts in the RGB/HB/AGB. SGB targets appear to cover both regions instead. The separation between the two groups of stars in indicated by the red line. A clear separation between first and second generation stars also appears, indicated by the black line. All stars with [N/Fe]$<$0.2 dex belong to the first generation (FG), while stars with [N/Fe]$>$0.2 dex belong to the second generation (SG).

At this point a possible concern is any systematic error that could affect our analysis since we used spectra with different resolutions and also different features for the nitrogen abundance determination. As far as C is concerned, we used the same feature (the G-band) although with a larger spectral interval for the low-resolution spectra of MS stars in order to have a better precision. To check a possible systematic error introduced by this choice we selected a sub-sample of 10 MS stars and determine their C content using the same narrower interval we applied to the other higher-resolutions data, i.e. 4300 $\AA$-4315 $\AA$. We found that the systematic error, if present, is lower than 0.05 dex and compatible with 0. If we consider N instead, we used a completely different CN feature for low resolution MS spectra because there only the UV CN feature was measurable, while in intermediate and high resolution spectra only the CN feature at 4215 $\AA$ was visible. Unfortunately we have no stars in common between the two datasets in order to verify the present of possible systematics. However in fig.\ref{fig4} we see that SG MS stars (the black points above the black line and on the right of the red line), whose N abundance was obtained using the UV CN feature, overlap nicely in their N content with SG SGB stars (the red points above the black line and on the right of the red line), whose N abundance was obtained using the CN feature at 4215 $\AA$. This indicates that any systematic error, if present, is small enough not to affect our results. 

\section{FG and SG stars in the CMD} \label{FSG}

We can now separate our targets into FG and SG stars and see where they are located on the CMD. For this purpose we indicated FG stars with red insets and SG stars with blue insets in figure \ref{fig2}. We can see immediately that, as far as the RGB is concerned, FG and SG stars are clearly separated, with almost no overlap, and with SG being bluer. The color difference is on average constant and of the order of 0.03 mag. AGB and HB targets present the same behavior, but with a certain overlap. SGB targets are totally mixed while the situation for the MS is better shown in the inset plot of figure \ref{fig2}. Even if FG and SG stars appear to be mixed, we can see that on average FG stars tend to follow closely the isochrone, while SG tend to be bluer, especially for V$>$18.4. This behavior will be explained in section \ref{sad}, but we can anticipate that it is a combination of selection effects, binarity and that SG stars are 0.03-0.04 mag bluer also in MS.

\begin{figure}
   \centering
   \includegraphics[width=\hsize]{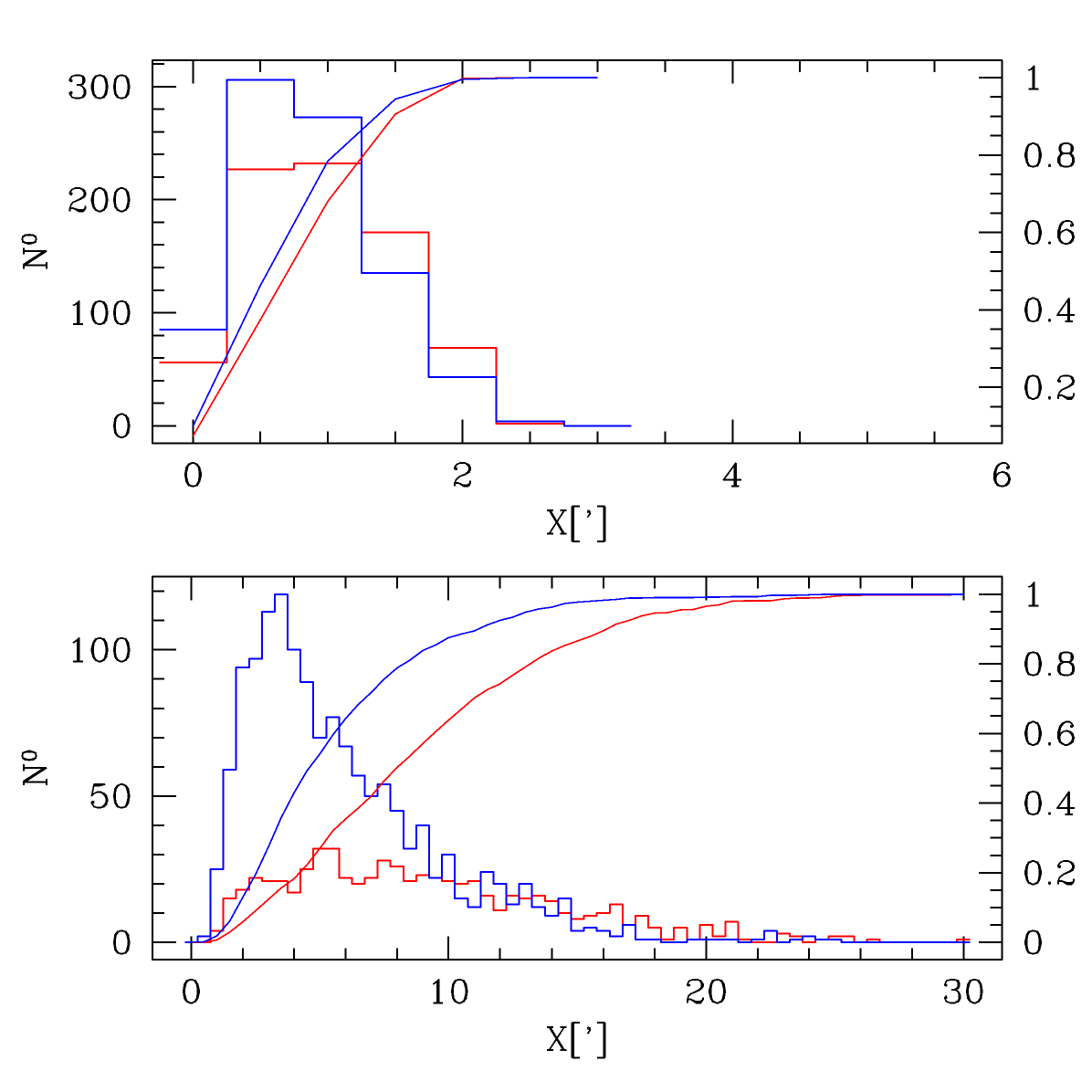}
   \caption{Histogram distribution and cumulative distribution of FG (red) and SG (blue) RGB stars. Histograms and cumulative curves from r=0'and r=3' were obtained from the HST photometry (upper panel), while those from r$\sim$2' to r=30' were obtained from the ground-based photometry (lower panel). Note that the x-scale is different for the two panels and that for the upper panel was extended to 6' in order to improve the visualization.}
    \label{fig8}
    \end{figure}

We asked ourselves what is the physical cause of this bimodality in the RGB and how it can affect also the rest of the CMD. The first suspect in such
cases is always differential reddening, although for this cluster this hypothesis seems to be very unlikely for two reasons. The first is the very low reddening (E(B-V)=0.03, \citealt{har10}) which would translate in a even lower (if not negligible) differential reddening ($\Delta$E(B-V)$_{\rm max}$=0.02, \citealt{pan24}). The second is that, if FG and SG are intrinsically mixed in the CMD (for example in the RGB phase), it would be extremely unlikely (if not impossible) for the differential reddening to separate them in a blue RGB, composed only by SG stars, and a in red RGB, containing only FG stars. 
In any case we decided to dig deeper into this issue. First of all, we selected blue and red RGB stars both in the ground-based and the HST photometries (see figure \ref{fig5}) and checked the spatial distribution of the two populations. 
In figure \ref{fig8} where we report the histogram distribution and cumulative distribution as a function of the radius of the two populations. Several things can be noticed:

\begin{enumerate}
    \item In the central part (within 3') SG slightly dominated, but it appears to drop faster than FG. In any case
    the cumulative distributions in this regions do no differ too much.
    \item Between 3' and  10' SG largely dominates.
    \item Beyond 10' FG starts to dominate and at 15' almost only FG stars are present.
    \item The cumulative distribution beyond 3' shows that SG is more concentrated.
    \item It is difficult to say anything about what happens at 3' since it correspond to the transition between the two photometries, the ground-based being largely incomplete for smaller radii.
    \item A Kolmogorov-Smirnov test indicates that, while in the central part (within 3') the difference between the two cumulative distributions is not significant (D$_{sample}$=0.10<D$_{ref}$=0.17 at a significance level of 1\%), in the outer part (beyond 3') the FG and SG populations really have a different distribution, with the latter being more concentrated (D$_{sample}$=0.30>D$_{ref}$=0.06 at a significance level of 1\%).
\end{enumerate}

This points against differential reddening being the cause for the CMD dichotomy. A final check is to ask what would happen if the 0.03 V-I color separation between the two RGB is caused by differential reddening and try and correct the CMD accordingly. A $\Delta(V-I)$=0.03 translate into a $\Delta(B-V)$=0.025, which translates into a $\Delta(V)$=0.07, incompatible with the SGB narrowness visible in figure \ref{fig2}, that has a V spread ($\sigma$) of 0.03 magnitudes. The two values are incompatible, ruling out the possibility for differential reddening to play an important role in shaping the CMD.

Since differential reddening is not the cause of the split we see in the RGB, we have to look toward other directions. One first candidate is the abundance difference in C,N,O and He affecting the two populations. According to our first results, the FG has [C/Fe]$\sim$-0.1 and [N/Fe]$\sim$-0.2 while 
the SG has [C/Fe]$\sim$-0.4 and [N/Fe]$\sim$+0.6. \citet{car09} finds [O/Fe] +0.3 and +0.1 for the two populations. If we calculate a synthetic spectrum for such abundances and typical RGB parameters (T$_{eff}$=5000 K, log(g)=2.50), and apply the V and I filter transmission curves given by the Asiago Database on Photometric Systems \citep{mor00} \footnote{http://ulisse.pd.astro.it/Astro/ADPS/}, we obtain a color difference that is almost negligible ($<<$0.01 mag).
We find the same result also if we apply an He variation for the second generation. In this last case we also took into account the effect of He enhancement calculating ad-hoc atmospheric models. Our simulation shows that, in order to have a color difference $>$0.01 mag, we have to use Y$>$0.40 for the second generation, a helium abundance never found in any globular cluster.

\begin{figure}
   \centering
   \includegraphics[width=\hsize]{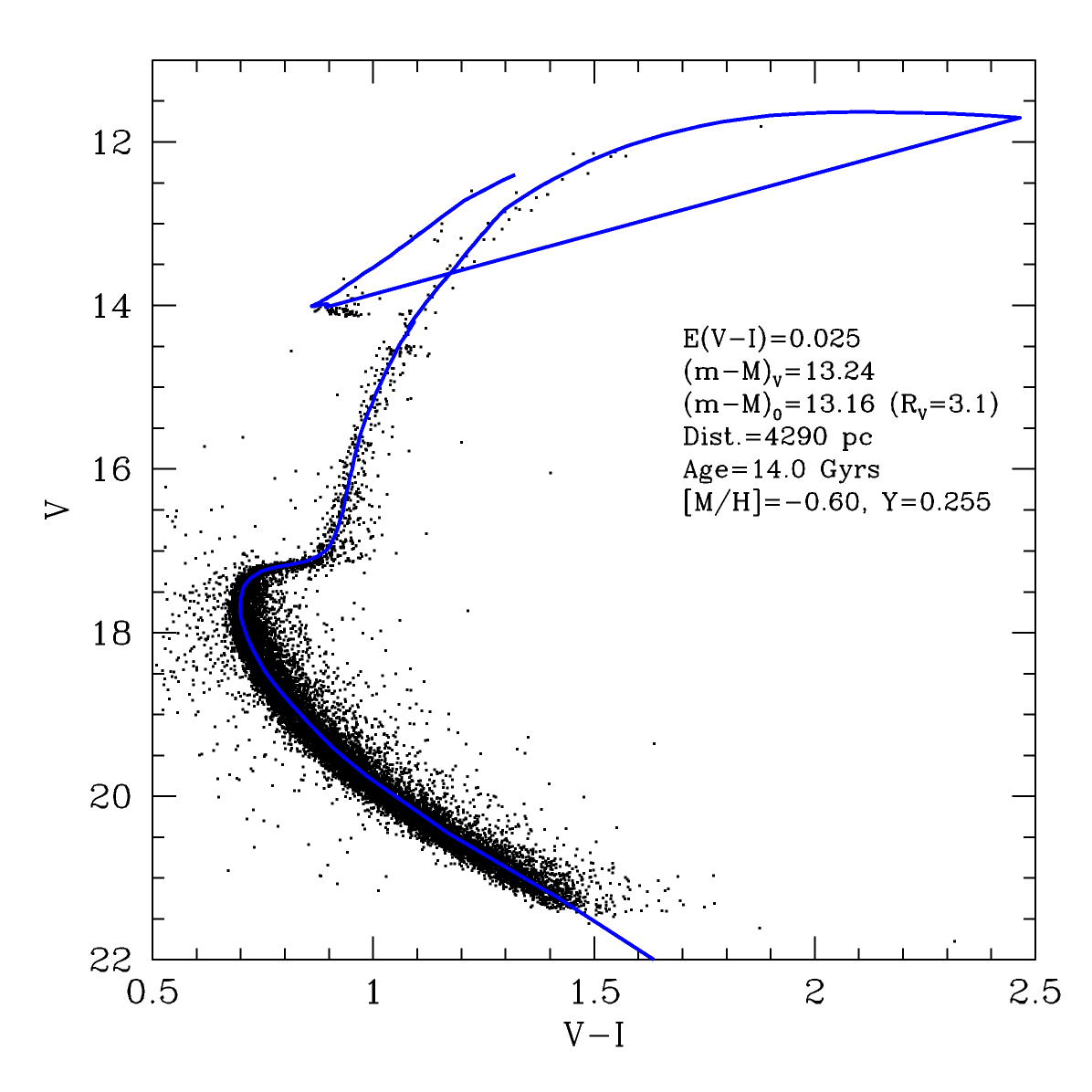}
   \caption{V vs. V-I CMD of the outer part of the cluster (r$>$15´). The blue line is the same  best fitting isochrone shown in figure \ref{fig2}. The parameters of the fitting are reported. See text for more details.}
   \label{fig9}
   \end{figure}

The only possibility left to explain the fact that RGB SG star are bluer is that they are intrinsically hotter. This can also be deduced from the \citet{alo99} paper that, as far as the T$_{\rm eff}$ vs. V-I relation for giants is concerned, finds no direct relation between the color and the metallicity of a star, just between color and temperature. That means that two giants with the same temperature always have the same V-I color, even if their metallicity is different. However the indirect cause for giant stars to be hotter is the global metal content [M/H] as can be inferred from isochrones. In fact, if we compare isochrones with different [M/H] values, the RGB part of that with the lower metallicity has higher temperatures and is bluer. For this reason the cause for SG to have a bluer RGB is that it has a lower global metallicty [M/H] or, in other words, a lower Z value. The way to prove this hypothesis is to perform a double isochrone fitting of the entire CMD. We start from the outer part where, as we have shown, we can isolate the FG. The CMD of stars having r$>$15' is reported in figure \ref{fig9}. As we can see, the same isochrone we used in figure \ref{fig2} is now an almost perfect fit of the data in all the evolutionary phases and all the discrepancies we mentioned in section \ref{fda} are no longer present. We obtained the following final parameters: (m-M)$_V$=13.24, E(V-I)=0.025, (m-M)$_0$=13.16 (distance=4290 pc, assuming R$_V$=3.1), Age=14.0 Gyrs and [M/H]=-0.60. We also assume Y=0.0255, that gives a good fit to the lower part of the RGB, which shape is sensible to the helium content. Typical systematic uncertainties are 0.10 and 0.01 mag on the distance modulus and reddening respectively, and 0.1 dex in [M/H]. The error on age is 1 Gyr.

\begin{figure}
   \centering
   \includegraphics[width=\hsize]{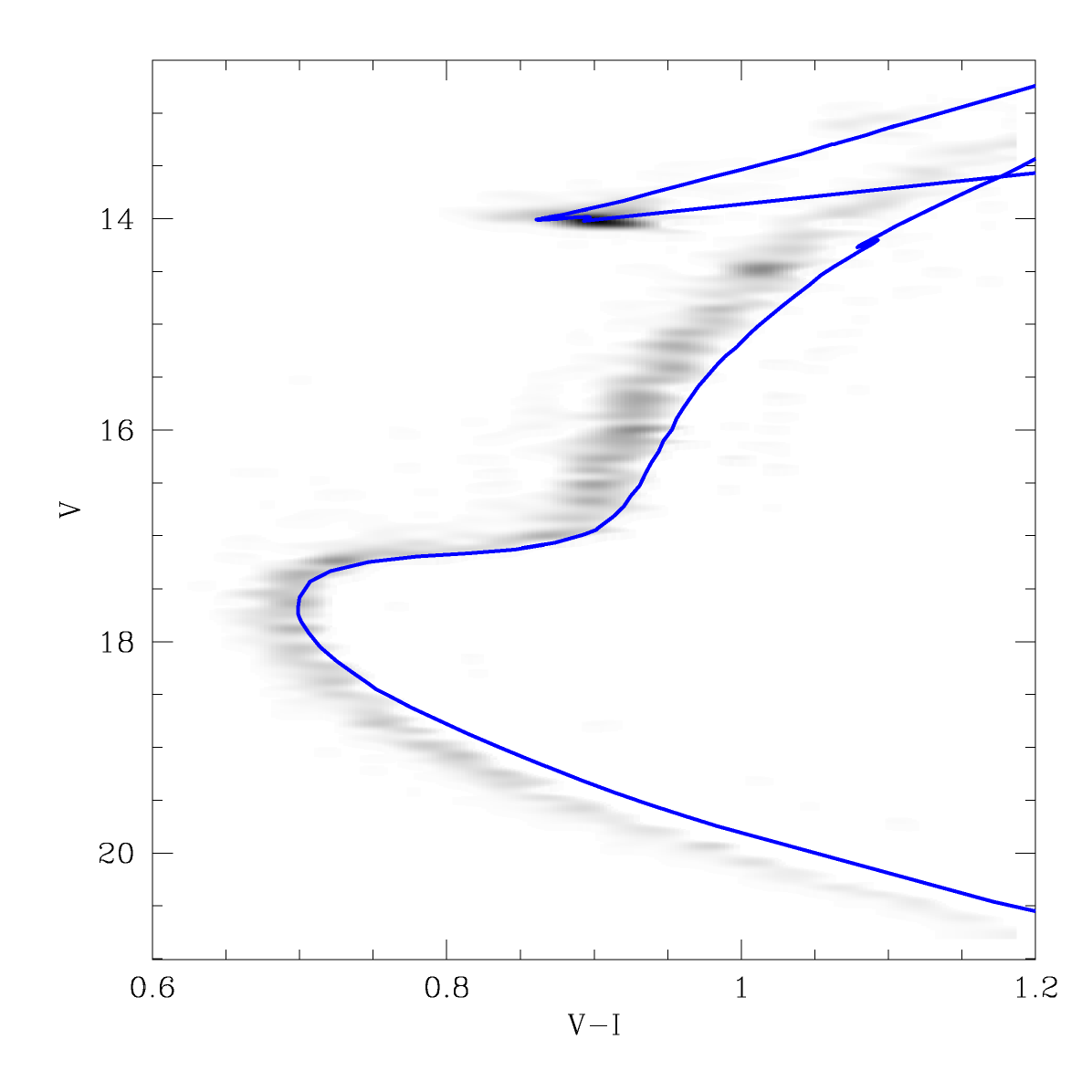}
   \caption{Difference between the Hess diagrams of the full field photometry (see figure \ref{fig2})
   and the outer part (r$>$15') photometry (see figure \ref{fig9}). This diagram enhances the position of SG stars in the CMD. The best fitting isochrone from figure\ref{fig2} is reported as reference.}
    \label{fig10}
    \end{figure}

If we move to the central part (r$<$15'), FG and SG are mixed together but we can highlight the position of the SG using the Hess diagram. For this purpose
we report in figure \ref{fig10} the difference of the Hess diagram of the full field photometry and the r$>$15' photometry. The isochrone of figure \ref{fig2} is also reported as reference. It is clear that the SG not only has a bluer RGB, but also a bluer TO and MS. Some caution must be used with the MS, however. This is because, while the SG RGB is well detached from its FG counterpart, the SG MS is partially merged with its FG counterpart, so that the MS we see in figure \ref{fig10} is just the blue wing of the total SG MS. This fact must be taken into account for the following isochrone fitting.

\begin{figure}
   \centering
   \includegraphics[width=\hsize]{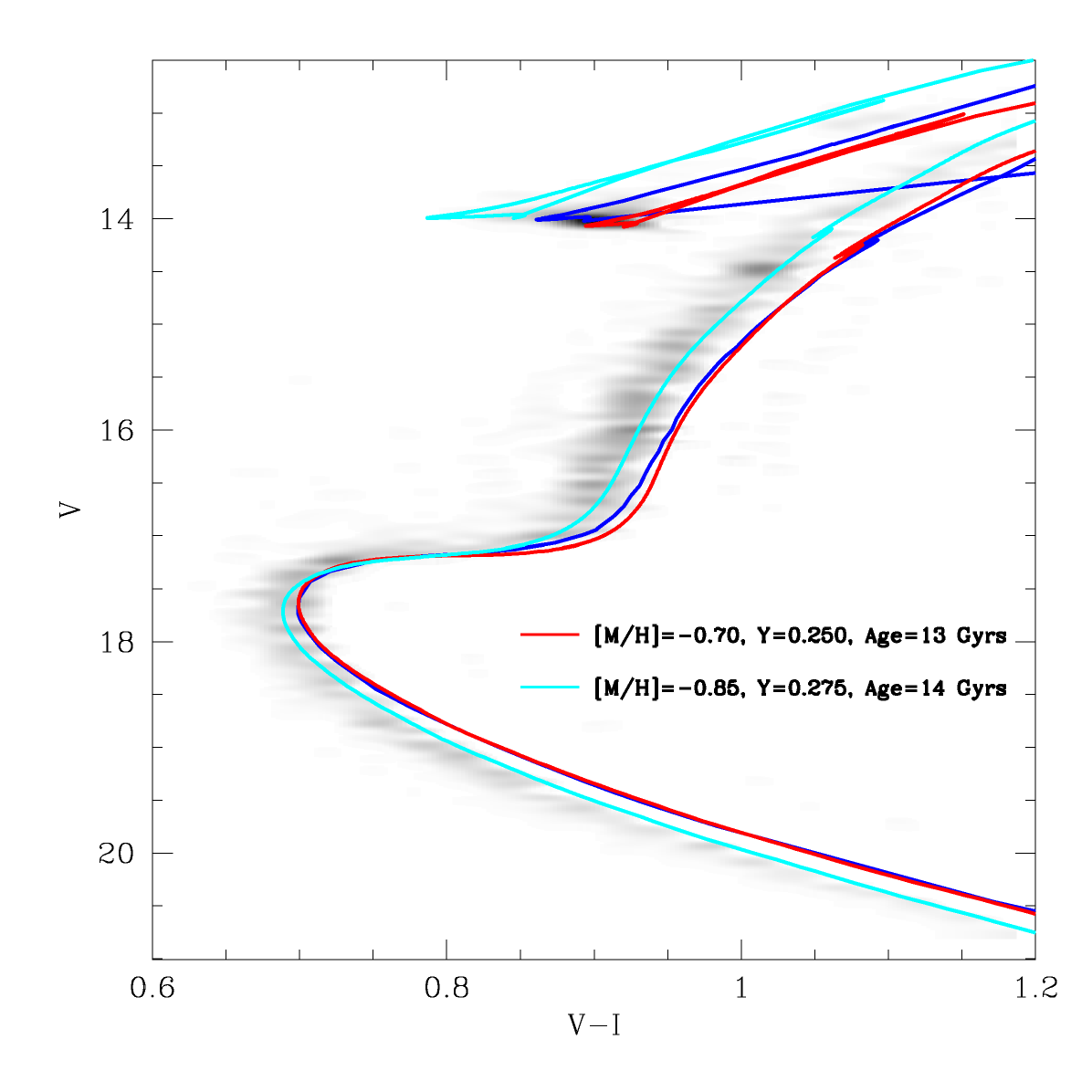}
   \caption{Difference between the Hess diagrams of the full photometry (see figure \ref{fig2})
   and the outer part (r$>$15') photometry (see figure \ref{fig9}). The blue line is the same  best fitting isochrone shown in figure \ref{fig2}, while the red curved is the BASTI isochrone that bast matches the Padova isochrone. Cyan curve is the He-enhanced, [M/H]-poor BASTI isochrone that best reproduce the locus of SG stars. Parameters for the two BASTI isochrones are indicated.}
    \label{fig11}
    \end{figure}

In order to obtain a proper CMD fitting, we have to take also He variation into account, especially to better reproduce the lower RGB and the MS. Unfortunately this is not possible with the isochrone database we used until now, so we are forced to move to the BASTI database \citep{pie21} \footnote{http://basti-iac.oa-abruzzo.inaf.it/index.html}, which gives the possibility to use different values for Y. First of all we selected the He normal (Y=0.25) isochrone that best reproduces the Padova model we used until now. This is the red curve in figure \ref{fig11}. The main differences are that for this new model we had to choose [M/H]=-0.70 (instead of [M/H]=-0.60) and an age of 13 Gyrs (instead of 14 Gyrs). This mismatch is not a surprise since different models usually assume different initial condition, metallicity scales and magnitude transformations. Having the BASTI isochrone that best fit the FG, we can now move to the SG. In order to reproduce its whole bluer RGB we have to choose a lower [M/H]. We found [M/H]=-0.85 as the best guess. This means that SG is significantly more metal poor than FG. As far as Y is concerned we got the best lower RGB (17<V<15.5) and MS fitting using Y=0.275. An higher value (Y=0.30 is available) would give a too blue MS. We have to mention also that, in order to fit properly the SG TO (18<V<17.5), we had to use an older age of 14 Gyrs, that means a SG 1 Gyrs older than the FG. If we use 13 Gyrs instead, we get a TO too blue. The differential error in the parameter determination between FG and SG is below 0.05 dex in [M/H], of the order of 0.01 in Y, and of the order of 1 Gyr in age. 

The resulting isochrone fit is reported in figure \ref{fig11}. We can see here that the metal-poor He-rich cyan isochrone reproduce very well the RGB and TO, but also the MS for which, as we mentioned before, we have only the blue wing. 
Our conclusion is that the V vs. V-I CMD of 47 Tuc cannot be fitted properly by just one isochrone, but two must be used, the main differences between the them being the global metallicity [M/H] (SG is 0.15 dex more metal-poor), He content (SG is 0.025 enhanced in Y) and possibly age (SG appears to be 1 Gyr older). The Y enhancement we found is consistent with the value reported by \citet[$\Delta$Y=0.03]{sal16}.

\begin{figure}
   \centering
   \includegraphics[width=\hsize]{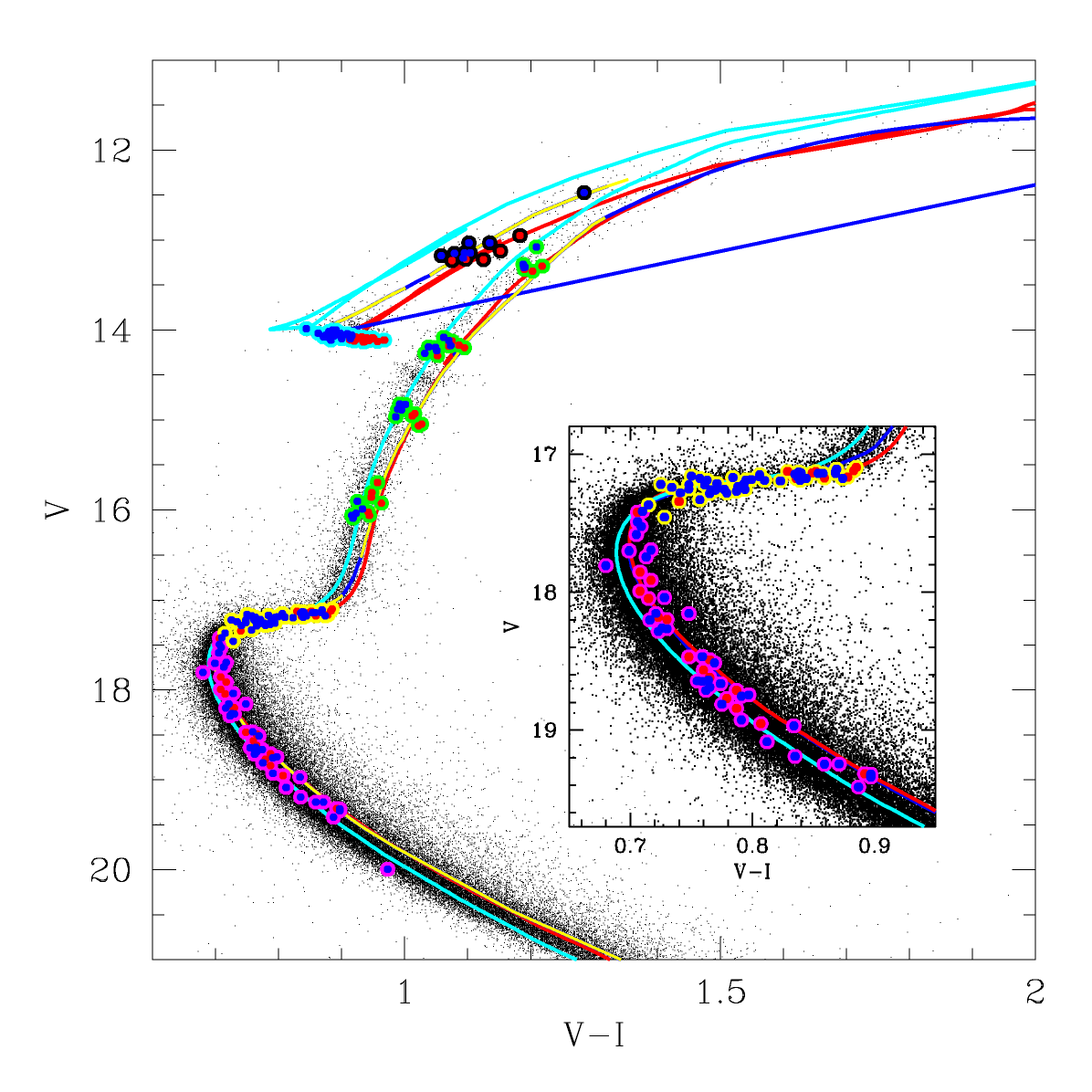}
   \caption{Same as figure \ref{fig1}, but with the best fitting Padova and BASTI isochrones for the FG and SG stars reported. Our targets are also reported. See text for more details. }
    \label{fig13}
    \end{figure}

We finally report the same CMD of figure \ref{fig2} in figure  \ref{fig13}, this time with the double-isochrone fit. We see how well the position of our FG and SG targets is reproduced, expecially in the RGB. As far as the MS is concerned, now also the bluer SG targets are reproduced by the cyan isochrone. If we move to the HB and AGB we see that FG and SG stars are still separated but not as well as in the RGB. We can infer that the cause for this behaviour is the differential mass-loss that can affect stars near the tip of the RGB. If two stars of the same population suffer a different mass-loss, they end up at different positions in the HB and follow different AGB path even if they have the same initial mass and chemical composition, being bluer that with the larger mass-loss. For this reason it is not surprising that the color spread of the two populations is larger in the HB than in the RGB, and that they overlap both in the HB and in the AGB phases. 

\section{Second abundance determination} \label{sad}

\begin{figure}
   \centering
   \includegraphics[width=\hsize]{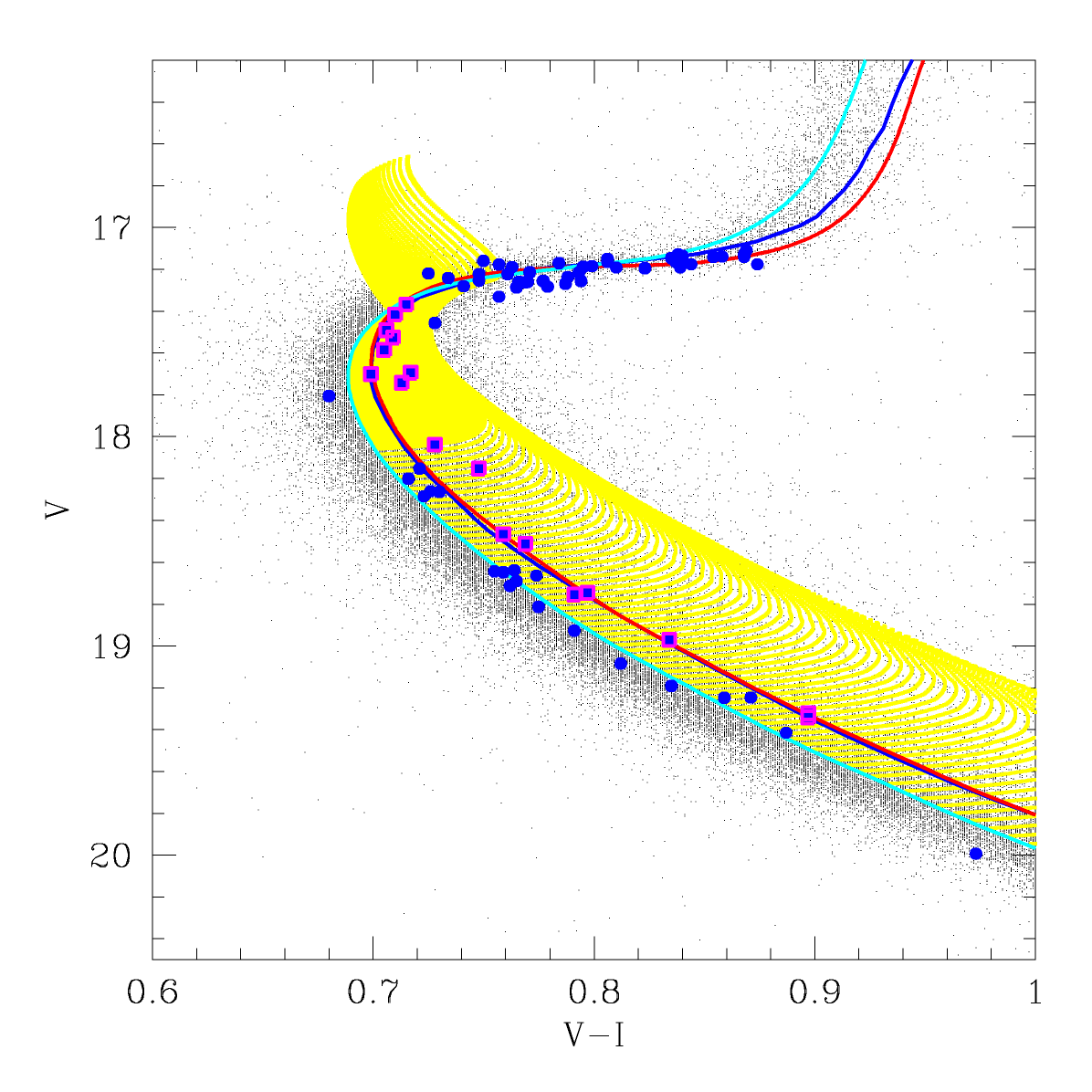}
   \caption{CMD position of our SG MS and SGB targets. Isochrones are those described in figure \ref{fig11}. Targets highlighted by a magenta square are those considered as binaries. See text for more details.}
    \label{fig14}
    \end{figure}

In the previous section we found that the only way to properly interpret 47Tuc CMD is by using two isochrones having different parameters. Now we want to discuss the impact of this result on the parameter and abundance determination of our targets. Parameters for HB and SGB stars were obtained from the V-I color which is sensitive only to temperature, so nothing changes for them. For AGB stars we highlighted an intrinsic color spread due to the fact that they belong to FG and SG and to a differential mass-loss during the previous RGB phase. For this reason the best temperature estimation for them is the T$_{\rm eff}$ vs. V-I relation that can be obtained from the corresponding portion of the isochrone (see figure \ref{fig1}) instead of the T$_{\rm eff}$ vs. V relation we used before. As far as RGB targets are concerned, FG and SG isochrones have a temperature difference of 55-65 K, with SG isochrone being hotter. For this reason we just adjusted temperatures for RGB SG targets of +60K. The situation for MS targets is a bit more complex. The inset in figure \ref{fig13} shows that FG MS targets follow nicely the FG isochrone, with some outlier that can be explained by photometric errors. SG targets instead show a different behavior. A fraction of them follow closely the SG isochrone, while the other fraction appears to follow the FG isochrone.
We investigate this weird behavior in figure \ref{fig14}, where we plot only SG targets. In this figure we build the SG binary sequence (the yellow region) by adding to each SG isochrone point all the fainter points using the equation of section 5 in \citet{alb21}. We see clearly that the targets follow either the metal-poor SG isochrone or the metal-rich FG isochrone. These last ones are indicated with magenta squares. Our explanation is that they are SG stars, but their redder position is due to the fact that they are in a binary system. So in the MS sample we have single FG stars and a mix of single and binary SG stars. The reason for that is probably the following. When \citet{bri04} selected the targets, they defined a mean ridge line for the cluster along the MS/TO/SGB region located likely in the middle of the two isochrones of figure \ref{fig14}. Then they include in their sample only stars within a certain distance from this mean line, and by this criterion they included only single FG stars but both single and binary SG stars. 
In any case, in order to have the best temperature estimation possible, we have to treat single and binary SG stars in a different way. For FG targets we stay with the same T$_{\rm eff}$ we obtained in section \ref{fda}. For single SG targets we modified the T$_{\rm eff}$ we obtained in section 
\ref{fda} by adding the mean temperature difference between the two isochones in their MS part, that is by adding +60K. For binary SG targets we just maintain the same T$_{\rm eff}$ without any change. In the following discussion we will check if this decision is the correct one. 
As far as gravity is concerned, we check that the best option was, for all the targets, to follow the same methodology explained in section \ref{fda} without any change. Finally microturbulence was obtained using the \citet{mot20} relation and the new temperatures. We report these final parameters in table \ref{table1}.

\begin{figure}
   \centering
   \includegraphics[width=\hsize]{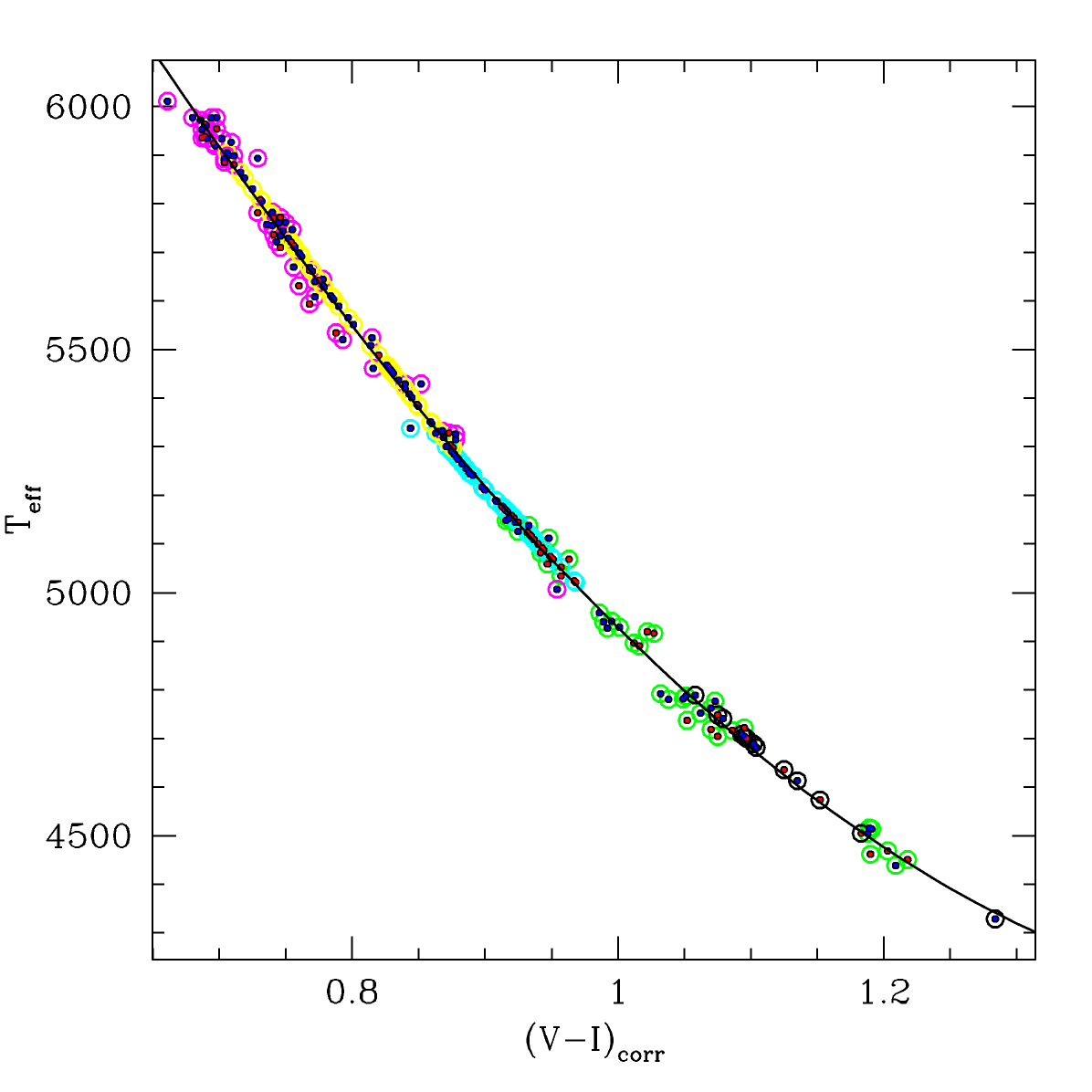}
   \caption{The T$_{\rm eff}$ vs. (V-I) relation obtained for our targets. V-I color of MS ans SGB stars was corrected for the gravity effect. The black line represents a 3$^{rd}$ degree polynomial fit. See text for more details.}
    \label{fig15}
    \end{figure}

At this point it is worth to check our parameters. First of all we report in figure \ref{fig15} our temperatures as a function of the V-I color. We have to
consider however that stars with higher gravities tend to have redder a V-I color for the same temperature. To correct for this effect we took a typical HB star as reference (T$_{eff}=5000K$, log(g)=2.5) and calculated its synthetic spectrum. We calculated then two spectra with the same temperature but with log(g)=4.0 (typical of a SGB target) and log(g)=4.5 (typical of a MS target). We found corrections of ${\rm -0.009}$ and ${\rm -0.019}$ mag respectively. The two corrections were applied to the colors of MS and SGB targets in figure \ref{fig15}. We can see that our targets define a nice and narrow sequence with a mean temperature spread around the polynomial fit of 22 K. 

\begin{figure}
   \centering
   \includegraphics[width=\hsize]{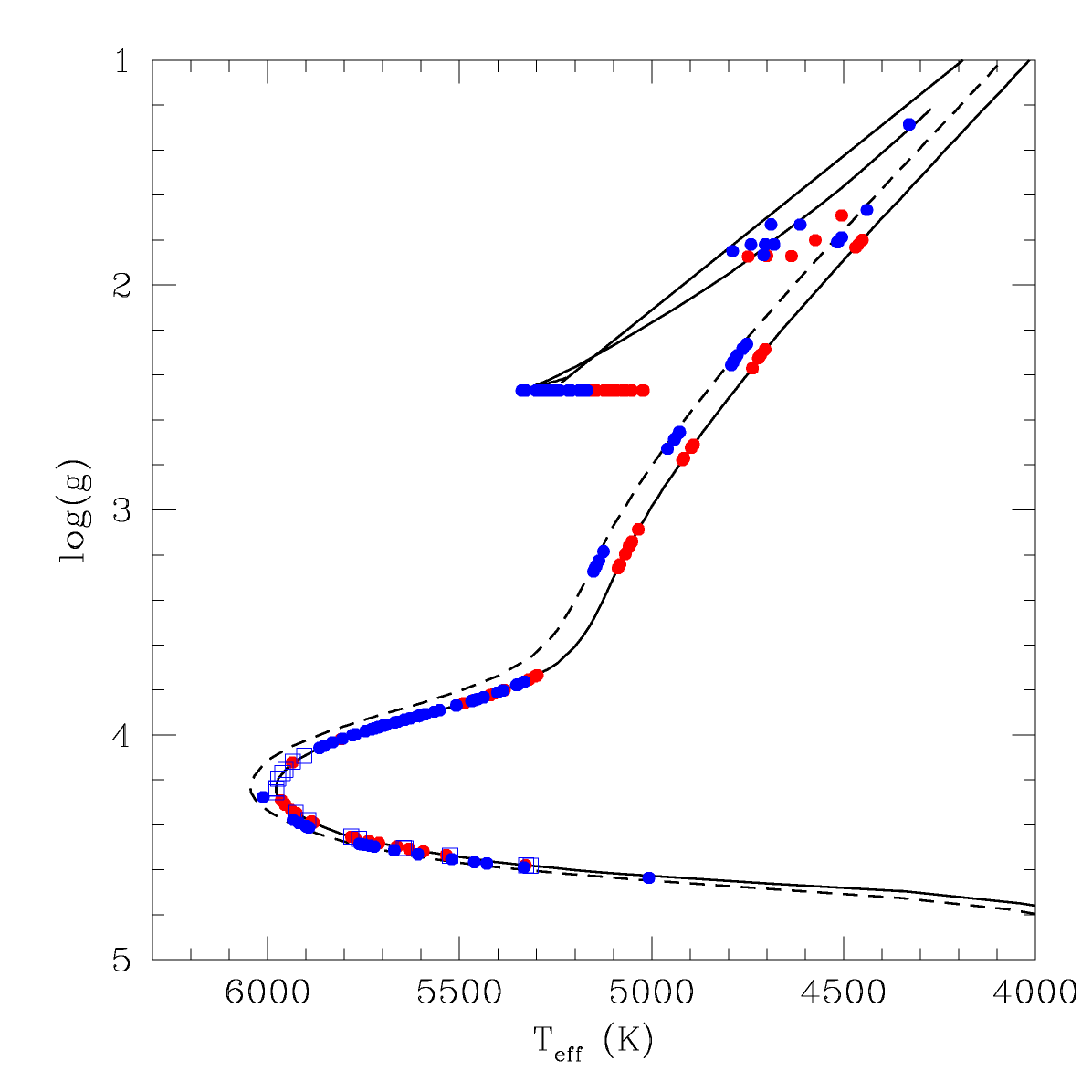}
   \caption{The log(g) vs. T$_{eff}$ for our targets. The continuos black lines represents the isochrones we used to obtain the initial stellar parameters (see fig \ref{fig2}). The shaded black lines is an isochrone from the same database but with a global metallicity 0.15 dex lower. Red circles are FG targets, while blue circles are SG targets. See text for more details.}
    \label{fig16}
    \end{figure}

We finally report in figure \ref{fig16} temperature and gravities and compare them with the isochrone of figure \ref{fig2}. We plotted also an isochrone from the same database but with a global metallicity 0.15 dex lower, as a reference. The spread around the first isochrone is due to the temperature corrections we discussed above, and the SG targets (which are 0.15 dex more metal poor that the FG) are well represented by the second isochrone at the MS and RGB level, where such corrections were applied. We conclude that no major internal systematic errors appears and that our parameters are consistent among all the evolutionary phases.

\begin{figure}
   \centering
   \includegraphics[width=\hsize]{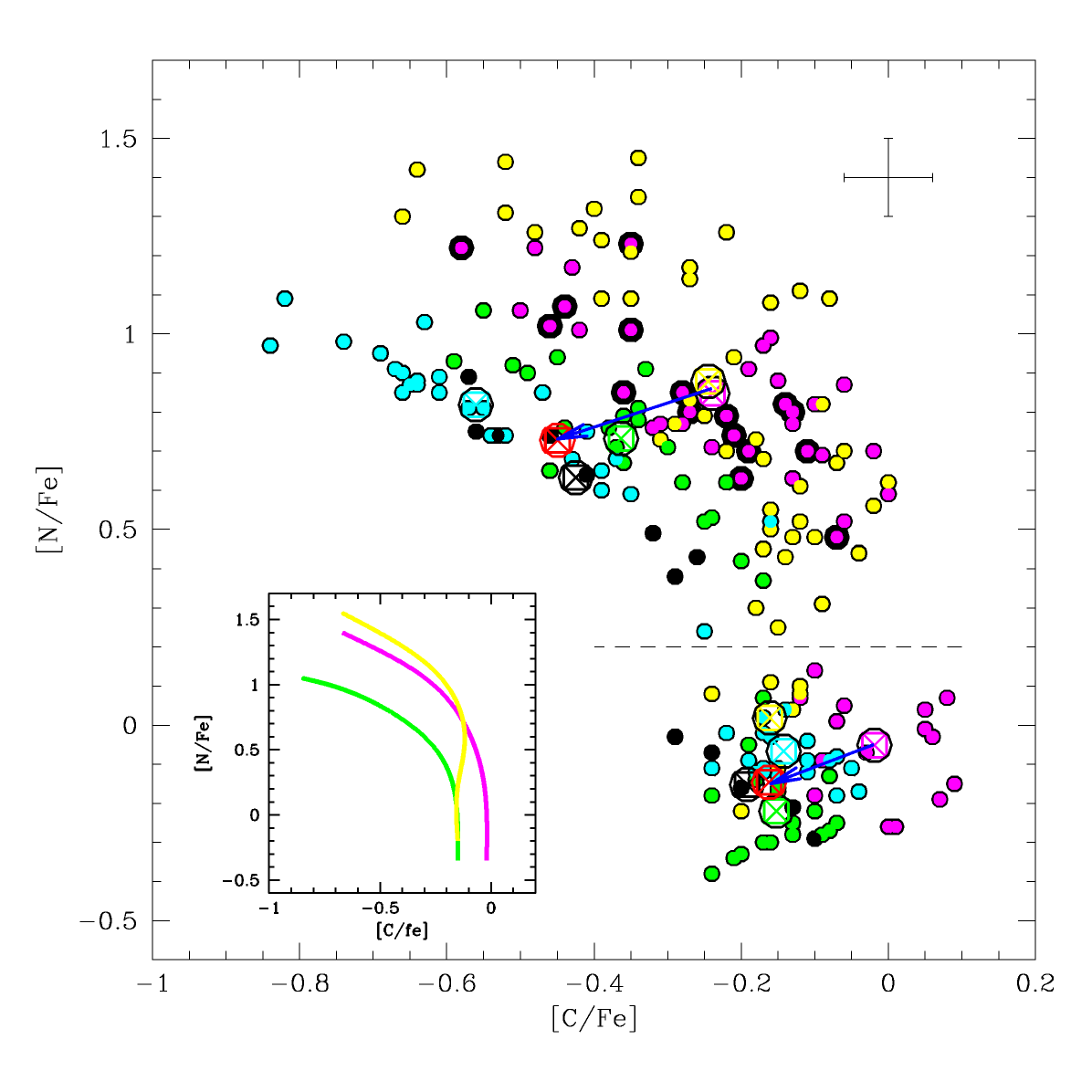}
   \caption{The final [N/Fe] vs [C/Fe] anticorrelation we obtained for our targets. Magenta, yellow, green, cyan and black symbols represents MS, SGB, RGB, HB and AGB targets respectively. Mean values for FG and SG stars are indicated with crosses surrounded by black circles. The horizontal line at [N/Fe]=+0.20 divides FG from SG targets. The inset shown the mean ridge lines for the N vs. C anticorrelations of MS (magenta), SGB (yellow) and evolved (green) targets. Error-bar shows the typical errors. See text for more details.}
   \label{fig17}
    \end{figure}

   \begin{figure}
   \centering
   \includegraphics[width=\hsize]{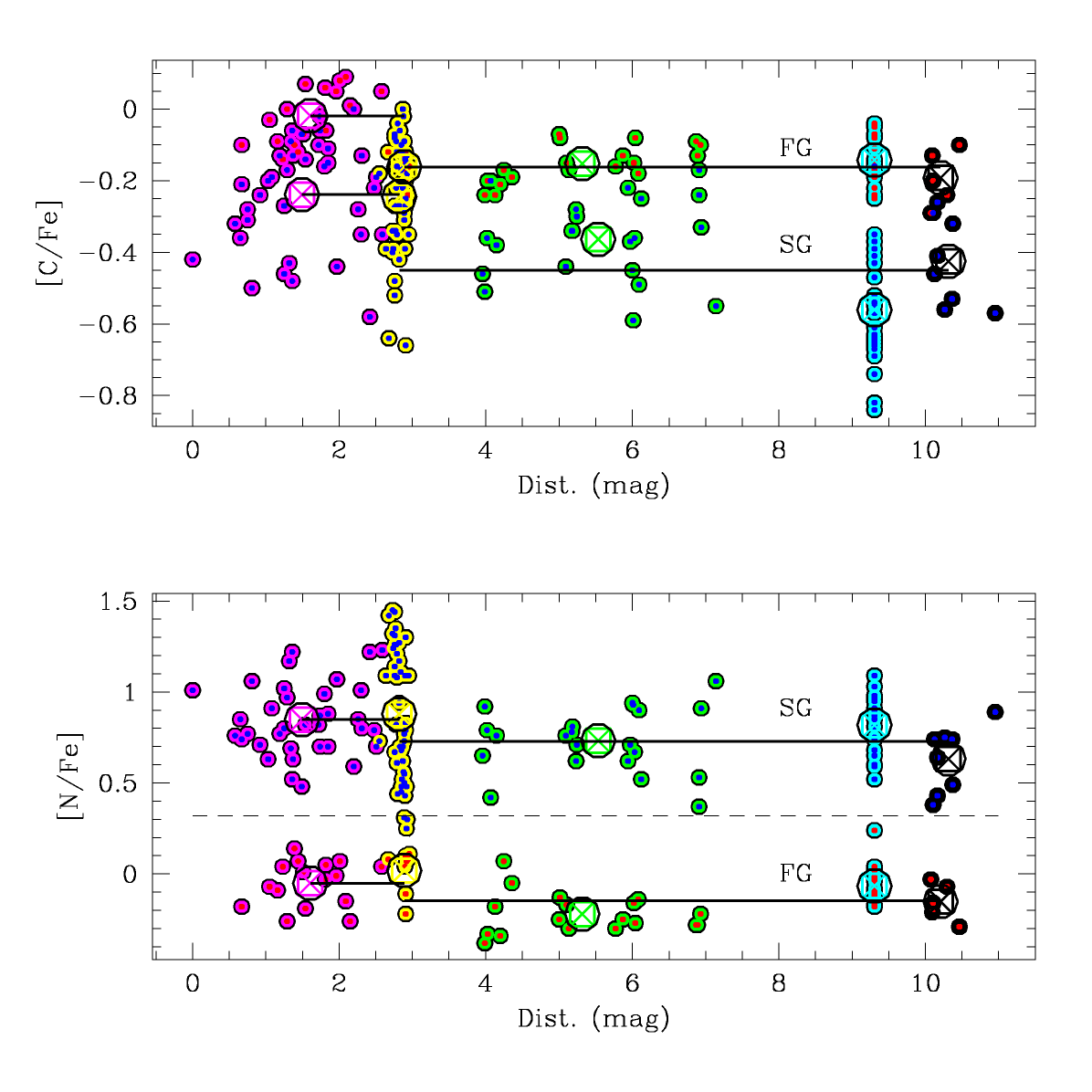}
   \caption{[C/Fe] vs Dist. (upper panel) and [N/Fe] vs. Dist. (lower panel) plots for our targets. Magenta, yellow, green, cyan and black symbols represents MS, SGB, RGB, HB and AGB targets respectively. The targets have a red or blue circle inside if they belong to the FG or SG respectively. Mean values for FG and SG stars at the different phases are indicated with crosses surrounded by big black circles. The horizontal line at [N/Fe]=+0.20 divides FG from SG targets, while the continuous black lines represent the mean values of unevolved (before the SGB) and evolved (after the SGB) stars. See text for more details.}
    \label{fig18}
    \end{figure}

Another important issue to discuss is the internal errors on [C/Fe] and [N/Fe]. Uncertainties on T$_{eff}$, log(g) and v$_t$ translate on errors in the abundance determination as well as the S/N of the data. In addition  we have to consider that the C content we measure also is affected by the uncertainties on [N/Fe], and the same happens for N that is affect by uncertainties on [C/Fe]. In order to estimate internal errors on [C/Fe] and [N/Fe] we first have to estimate error on stellar parameters. From what we said before we can adopt 30K as a reasonable estimation for the internal error in temperature. Also, the isochrone fitting procedure suggests that 0.05 is a good value for the uncertainty on gravity, while for the microturbulence we adopt 0.05 km/s. At this point we selected one representative stars for each evolutionary phase and estimated new [C/Fe] and [N/Fe] by varying stellar parameters of the indicated amount one by one. The variation of the abundances is indicated in table \ref{table1B}, columns 2 to 5. Columns 6 gives the total error due to stellar parameters and S/N calculated as the root-square of the sum of the squares of the previous 4 columns. We now can calculate the error on [C/Fe] due to the error on [N/Fe] (column 7), and the error on [N/Fe] due to the error on [C/Fe] (column 8). Finally, column 9 gives the total error calculated as the root-square of the sum of the squares of the previous 3 columns. 

As we can see, the major contributors to the final errors are uncertainty on temperature and the S/N of the data. Also uncertainty on [C/Fe] has a significative impact on the final [N/Fe] abundance, while uncertainty on [N/Fe] does not affect [C/Fe]. Table \ref{table1B} shows that we can assume 0.06 dex as the typical internal error for [C/Fe] and 0.10 dex as the typical internal error for [N/Fe].
    
\section{Results} \label{res}

The new parameters were used to obtain the final abundances that are reported in figure \ref{fig17}, \ref{fig18} and table \ref{table1}. The main change with respect to figure \ref{fig4} is in the C abundances of single MS, SG targets. Now these abundances better overlap with those of the other MS targets and the MS's star N vs. C anticorrelation is significantly narrower. About the other targets, abundances did not change significantly. 

In figure \ref{fig17}, we report [N/Fe] vs. [C/Fe], i.e. the N vs C anticorrelation, together with the mean [C/Fe] and [N/Fe] values of each evolutionary phase with colored crosses surrounded by a black circles. These values are also reported in table \ref{table2}. Red crosses surrounded by red circles represent the mean C and N abundances for evolved FG and SG targets, i.e. the mean of the RGB/HB/AGB phases (see below for the reason why made this choice). We also reported the typical errors determined in the previous section. We can immediately see that the spread of the data is much larger that the typical error, making us confident that what we see in this figure is real. 
A value of [N/Fe]=+0.20 can be safely used to divide FG and SG targets as indicated by the dashed horizontal line of figure \ref{fig17} and this classification is reported in the column 12 of table \ref{table1}. While the observed spread of FG stars is comparable with the typical error, the spread of SG stars is much bigger. We can infer then that FG stars have an homogeneous C and N content, while SG stars have a real intrinsic C and N spread.

As far as FG is concerned, we see that [C/Fe] decreases of $\sim$0.14 dex moving from the MS to the later phases, while [N/Fe] decreases of 0.10 dex. For SG instead, [C/Fe] and [N/Fe] have the same value for MS and SGB, then [C/Fe] decreases of $\sim$0.21 dex moving to later phases, while [N/Fe] decreases of 0.12 dex. These trends are indicated by the blue arrows. The behavior of FG and SG SGB stars will be further discussed in the next paragraph.
We can also see that SG targets suspected to be binaries (the magenta circles surrounded by thick black circles) follow nicely the trend defined by non-binary targets, suggesting that no systematic shift is introduced by the methodology we adopted above in order to obtain the final atmospheric parameters.

There are a couple of mismatches that we have to underline. The first concerns those N-rich SGB stars with [N/Fe]>1.0. These targets have N abundances slightly higher than their MS SG counterparts. This is hardly an evolutionary effect and it could be due to some residual systematic error on the determination of the N content. In any can this will not affect the following discussion. The second concerns the N vs. C anticorrelation of RGB stars. While the FG RGB targets are well mixed with FG HBs and AGBs, the SG RGBs have a slightly higher C and N content compared to SG HBs and AGBs. This could be an evolutionary effect where, at the level of the RGB, the transition from MS to HB/AGB abundances is already completed for the FG, while it is still an ongoing process for the SG. In any case we will treat the RGB, HB and ABG phases all together.

The inset in figure \ref{fig17} shows the mean ridge lines for the N vs C anticorrelations of MS (magenta), SGB (yellow) and evolved (green) targets. While anticorrelations for MS and evolved targets are well detached, that of the SGB targets matches the FG part of the evolved anticorrelation, and then it moves gradually to match the SG MS anticorrelation. This behavior is due to two reasons. In the first place, during the SGB phase, stars change their superficial [C/Fe] and [N/Fe] abundances due to the first dredge-up, as we will discuss below. We call the SGB region where this effect that place {\it abundance changing region}. The other reason is due to selection effects. Infact, FG SGB targets and the most N-poor of the SG SGB targets are located after the {\it abundance changing region} (as we will show later) and so they follow the evolved anticorrelation. On the other hand, the most N-rich of the SG SGB targets are located before this region and so they follow the MS anticorrelation. 

Another thing we have to notice in figure \ref{fig17} is that, while
the mean FG SGB [C/Fe] abundance matches that of the FG evolved stars, the [N/Fe] abundances for the same group is higher instead. This is because the most N-poor FG SGB star are missed from the sample (i.e. there are very few yellow targets with [N/Fe]<0.0 in figure \ref{fig17}), very likely due to selection effects.
On the other hand the SGB SGB population is well represented and
its mean [C/Fe] and [N/Fe] abundances match those of the SG MS stars. 

Figure \ref{fig17} also shows that, if we compare HB targets (cyan) with the RGB (green) and the AGB (black), a tail of C-poor stars is missing. The explanation for this behavior is that, while we observe the entire HB and so we mapped the entire N vs. C anticorrelation down to the lowest C abundance, for the RGB and the AGB  we missed the most extremes targets, i.e. the most C-poor. In fact, these targets should be the bluest and we cut them off during the selection process. If we look carefully figure \ref{fig13}, we see actually that on the blue side of our RGB and AGB targets there are several stars (all members) and we can guess that those would populate the very C-poor tail of the N vs. C anticorrelation. The missing targets belong to the SG, as also clearly visible in figure \ref{fig17}. On the other hand figure \ref{fig17} shows that SG RGB and AGB targets map better the anticorrelation for evolved stars between [N/Fe]=+0.20 and [N/Fe]=+0.50.
So the best estimation for the mean C and N abundances of the evolved SG targets is the mean of the abundances of the SG RGB/HB/AGB phases. We adopted the same procedure to calculate the mean abundances of the FG evolved targets. These values are those reported as red symbols in figure \ref{fig17} and will be used in the following discussion.

Figure \ref{fig18} shows the trend of [C/Fe] and [N/Fe] with the evolutionary phase, that is represented by the distance (in magnitudes) along the isochrone from a MS reference point and the projection of each target onto the isochrone. In order to calculate the distance we followed this procedure:

\begin{itemize}
  \item Since the points of the isochrone are not evenly spaced, we interpolated it with a 0.01 mag. step.
  \item Then we associated to each target the closest interpolated point.
  \item Choosing the point closest to the faintest target (the MS star with V=20 in figure \ref{fig2}) as reference, we calculate the distance of each other point along the isochrone path.
  \item We did not consider the jump from the RGB-tip to the zero-age HB in our calculation, so a start on the RGB-tip and one on the zero-age HB have the same distance.
  \item For this procedure we used the isochrone from the Padova database (the blue line in figure \ref{fig13} and figure \ref{fig2}).
\end{itemize}

For each sub-population (MS,SGB,RGB,HB and AGB) we divided the targets in FG and SG groups. For each of them we then calculated the mean [C/Fe] and [N/Fe] abundances (the same reported in table \ref{table2}) and the mean distance. The mean values are reported as big black circles with inside a cross of the color of the sub-population.

\begin{table}
\caption{Mean [C/Fe] and [N/Fe] abundances for FG and SG for the different 
evolutionary phases.}             
\label{table2}      
\begin{tabular}{l c c c c}        
\hline\hline                 
Phase & [C/Fe]$_{FG}$ & [N/Fe]$_{FG}$ & [C/Fe]$_{SG}$ & [N/Fe]$_{SG}$\\   
\hline                        
 MS & -0.02 & -0.05 & -0.24 & +0.85 \\      
SGB & -0.16 & +0.02 & -0.24 & +0.88\\
RGB & -0.15 & -0.22 & -0.36 & +0.73\\
 HB & -0.14 & -0.07 & -0.56 & +0.82\\
AGB & -0.19 & -0.15 & -0.43 & +0.63\\ 
\hline                                   
\end{tabular}
\end{table}

In figure \ref{fig18} the mean abundance trends are indicated by black lines, each one named with the population they represent (FG or SG). As far as FG is concern, MS stars have the mean [C/Fe] abundance of -0.02 dex, that we assume remains constant until the SGB {\it abundance changing region} (we have too few FG SGB target before this region to be certain). After this region the SGB mean FG C abundance decrease to -0.16 dex and stays constant until the AGB. 
For the SG instead, MS stars have the mean [C/Fe] abundance of -0.24 dex that stays constant until the SGB {\it abundance changing region}. In this case a significant fraction of our targets lie before this region so we can be certain. Then it decreases to a value of about -0.45 dex and remains constant. 
The lower panel shows that also N abundances does have the same behavior, but with a much smaller decrease of -0.10 dex for the FG, and -0.12 dex for the SG. We can also say that we find no abundance variation at the RGB-bump level, that have log(g)$\sim$2.5 (Dist.=5.5). 

A decrease of $\sim$-0.15$\div$-0.20 dex in C after  the {\it abundance changing region} is compatible with the effect of the first dredge-up. \citet{vin21}  performed a test of theoretical stellar models that predicts a C-depletion of -0.14 dex after the first dredge-up, very close to the value we find. However the same models predict a N enhancement of +0.19 dex, while we find a N depletion of $\sim$-0.10$\div$-0.12 dex.
On the other hand \cite[see their figure 1]{sal20} predicts that, for an old population (13 Gyrs), the $\Delta$[N/Fe] expected is of the order of about -0.1 dex, close to the value we obtain. In any case the first dredge-up appear to the the cause of the C and N behavior we find at the level of  the {\it abundance changing region}. 

\begin{figure}
   \centering
   \includegraphics[width=\hsize]{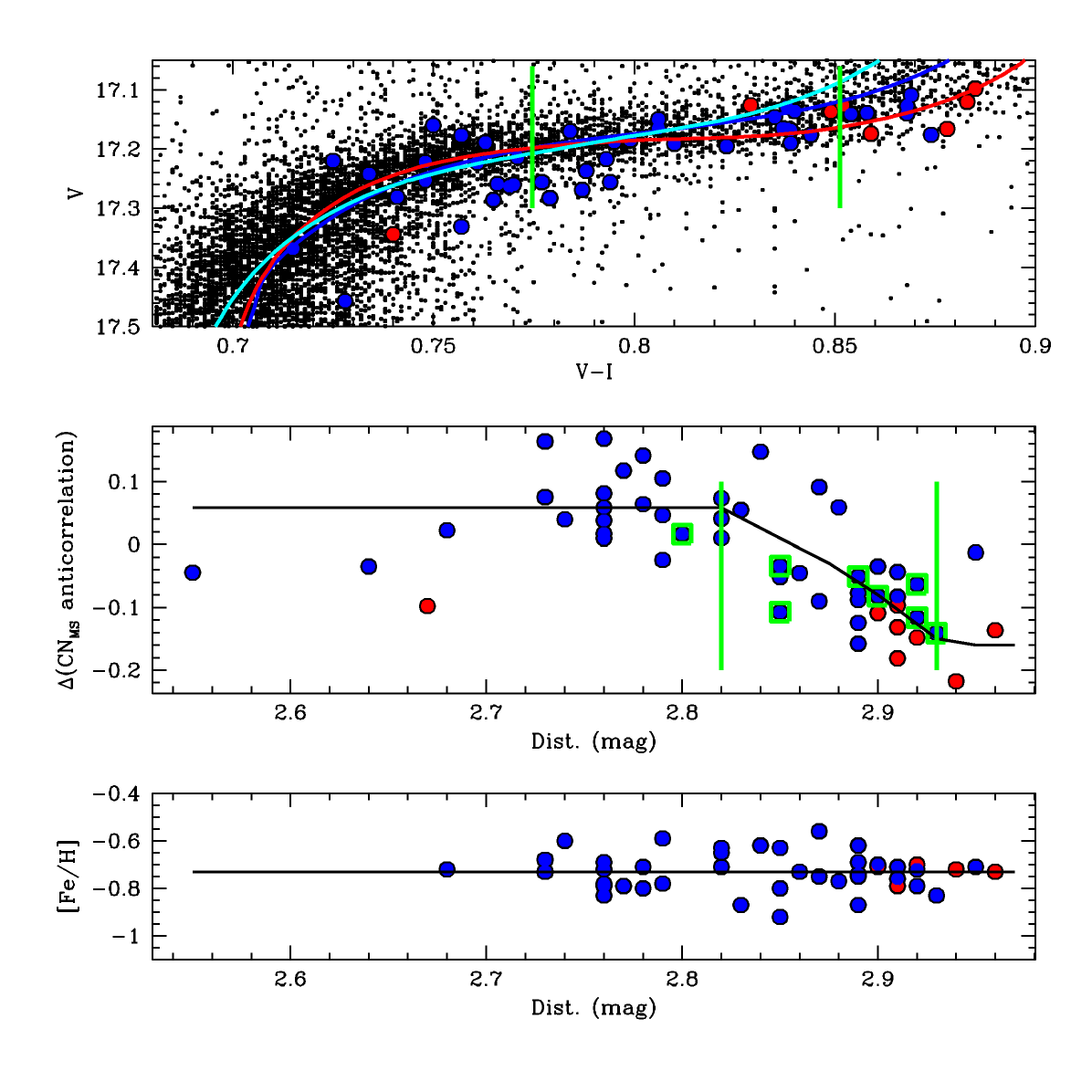}
   \caption{Upper panel: Distribution on the CMD of FG (red) and SG (blue) SGB targets. The SGB region where abundance change happens is indicated by the two green lines. Middle panel: Abundance variation along the SGB. The y coordinate is the [C/Fe] difference between each target and the point of the MS anticorrelation (the magenta line in the inset of figure \ref{fig17}) closest to that target. Bottom panel: [Fe/H] as a function of the position along the SGB. See text for more details.}
    \label{fig19}
    \end{figure}

Figure \ref{fig17} and \ref{fig18} show a clear decrease in carbon an nitrogen when stars move away from the MS, that is complete (or almost complete) at the RGB level, at least for V<16.0.
We want to investigate if we can exactly identify the region of the SGB were this decrease happens, that we previously called the {\it abundance changing region}. For this purpose we plot in figure \ref{fig19} the SGB region (upper panel)
with FG SGB targets in red and SG SGB targets in blue. In the middle panel we report in the X axis the distance parameters of the targets from the reference point (see above for the definition) and in the Y axis a parameter we call $\Delta$(CN$_{MS}$ anticorrelation). This parameter is the 
$\Delta$[C/Fe] difference between the target [C/Fe] and the point of the MS anticorrelation (the magenta line in the inset of figure \ref{fig17}) closest to that target. For this purpose we interpolated the MS anticorrelation with a 0.01 dex step. This panel shows clearly that both subpopulation show the same behaviour with 
the $\Delta$(CN$_{MS}$ anticorrelation) parameters that starts to decrease at Dist.=2.82, with the decreasing being complete for Dist.=2.93. These two limits are indicated by the two green lines. This region can be transformed back to the CMD and is the SGB part between the two green lines of the upper panel of figure \ref{fig19} (between V-I=0.77 and V-I=0.85). The cause of this transition is very likely the first dredge-up as we discussed before. In the same figure (bottom panel) we report also the [Fe/H] abundance of stars from \citet{mar16} (we do not have iron abundance for the other targets) and we see that iron abundance does not vary along the SGB. So we can safely assume that the drop in C and N is not accompanied by a drop in iron and that the variation in [C/Fe] and [N/Fe] are not due to a variation in [Fe/H]. The mean [Fe/H] for these stars is -0.73$\pm$0.01, very close to the value we assumed for all the targets ([Fe/H]=-0.70). 

We see also that almost all the FG SGB stars are located at the very end of the {\it abundance changing region}, when both C and N abundances reach their minimum value, while at least half of the SG SGB stars are located on the left side still keeping their MS C and N content. This can explain the weird behavior of the C vs. N anticorrelation of SGB targets visible in figure \ref{fig17}. Most of those that belong to the FG already suffered a decrease in C and N, so their trend matches that of the more evolved phases. On the other hand many of those that belong to the SG still retain their original MS C and N content and so their trend matches that of the MS. This is confirmed by the blue circles surrounded by green squares in figure \ref{fig19} (middle panel). These are the SG SGB target with [N/Fe] in the range 0.20$\div$0.50 dex and that in figure \ref{fig17} are the SG SGB targets that are on average the ones that most moved away from the mean MS anticorrelation. As expected, also these target are located toward the end of the SGB {\it abundance changing region} and they moved away from the MS trend because that already experimented the C and N abundance decrease.

Another question rises at this point. Why almost all FG SGB targets are located in the very red part of the SGB, while the SG SGB targets are spread all over? It hard to give an answer but a possible explanation is that FG SGB stars follow a different path along this CMD region (\citealt{and09} found a split here) and they were missed due to selection effects. Also, if we look carefully at figure \ref{fig17}, we see that FG SGB targets are almost all located at higher N abundances ([N/Fe]>0.0) compared to their evolved counterparts (RGB, HB and AGB) and that there is just 1 FG SGB star with a low N content ([N/Fe]=-0.20). This reinforces the idea that the low N-content FG SGB stars were completely missed during the target selection because they are located in a different part of the CMD, very likely at lower V magnitudes. Future investigations will clarify this point.

\section{Conclusions} \label{con}

In this paper we analyzed the V vs V-I CMD of the globular cluster NGC 104 both in the inner part and out to 25' from the center. We also analyzed a sample of low, medium and high resolution spectra covering CH and CN blue and UV bands in order to obtain [C/Fe] and [N/Fe] abundances for a sample of targets covering all the evolutionary phases, from the MS to the AGB. We found that:

   \begin{enumerate}
      \item Spectroscopic targets define a [N/Fe] vs. [C/Fe] anticorrelation where FG and SG stars can be easily separated.
      \item In the CMD, FG and SG targets define two distinct sequences that can be traced from the MS to the AGB.
      \item The SG is more centrally concentrated while the FG dominates beyond 15'.
      \item In order to fit properly the CMD, two isochrones are required. 
      \item The FG CMD can be fitted with an isochrone having E(V-I)=0.025, (m-M)$_V$=13.24, [M/H]=-0.70, Y=0.25 and Age=13.0 Gyrs.
      \item The SG CMD can be fitted with an isochrone having the same reddening and distance modulus, but with [M/H]=-0.85, Y=0.275. An older age of 14.0 Gyrs is also required to match the SG TO color.
      \item By consequence, the SG is more metal-poor, He-richer and possibly older than the FG.
      \item Both C and N decrease moving away from the MS. The decrease in C is more significant (-0.14 for FG and -0.21 for SG), while for N is of the order of 0.10 dex. 
      \item This abundance decrease happens at the level of the SGB, between V-I=0.77 and V-I=0.85 and it is very likely connected with the first dredge-up, both for the location where it occurs and for the amount of the decrease.
      \item The decrease in C and N is not accompanied by a drop in [Fe/H], that is constant along the SGB.        
   \end{enumerate}
\ \\
\ \\
\ \\
Table \ref{table1} is only available in electronic form at the CDS via anonymous ftp to cdsarc.u-strasbg.fr (130.79.128.5) or via http://cdsweb.u-strasbg.fr/cgi-bin/qcat?J/A+A/
\begin{acknowledgements}
S.V. gratefully acknowledges the support provided by Fondecyt Regular n. 1220264 and by the ANID BASAL project FB210003. The authors acknowledge the anonymous referee for the contructive report.
This work has made use of data from the European Space Agency (ESA) mission
{\it Gaia} (\url{https://www.cosmos.esa.int/gaia}), processed by the {\it Gaia}
Data Processing and Analysis Consortium (DPAC,
\url{https://www.cosmos.esa.int/web/gaia/dpac/consortium}). Funding for the DPAC
has been provided by national institutions, in particular the institutions
participating in the {\it Gaia} Multilateral Agreement.
\end{acknowledgements}

\end{document}